\documentclass[useAMS,usenatbib]{mn2e}

\usepackage{graphicx}
\usepackage{txfonts}

\title[Chemodynamical and photoionization modelling of galaxies]
      {Photoionization analysis of chemo-dynamical dwarf galaxies simulations}

   \date{Accepted for publication in Monthly Notices of the Royal Astronomical Society Main Journal, 12 March 2015.}

\author[Melekh et al.]
       {B.~Melekh$^{1}$, 
        S.~Recchi$^{2}$, 
        G.~Hensler$^{2}$, 
        O.~Buhajenko$^{1}$,\\ 
     $^{1}$ Department of Astrophysics, Ivan Franko National University,
            Kyrylo \& Methodiy str. 8, 79005 Lviv, Ukraine \\
     $^{2}$ Institute for Astrophysics, University of Vienna,
            T\"urkenschanzstrasse 17, A-1180 Vienna, Austria \\
       }

\def\Ha{{{\rm H}$\alpha$} }

\begin{document}

\maketitle

\begin{abstract}
Photoionization modelling allows to follow the transport, the
emergence, and the absorption of photons taking into account all
important processes in nebular plasmas.  Such modelling needs the
spatial distribution of density, chemical abundances and temperature,
that can be provided by chemo-dynamical simulations (ChDS) of dwarf
galaxies.  We perform multicomponent photoionization modelling (MPhM)
of the ionized gas using 2-D ChDSs of dwarf galaxies.  We calculate
emissivity maps for important nebular emission lines. Their
intensities are used to derive the chemical abundance of oxygen by the
so-called $T_e-$ and $R_{23}-$methods.  Some disagreements are found
between oxygen abundances calculated with these methods and the ones
coming from the ChDSs.  We investigate the fraction of ionizing
radiation emitted in the star-forming region which is able to leak out
the galaxy.  The time- and direction-averaged escape fraction in our
simulation is 0.35--0.4.  Finally, we have calculated the total \Ha
luminosity of our model galaxy using Kennicutt's calibration to derive
the star-formation rate.  This value has been compared to the 'true'
rate in the ChDSs.  The \Ha-based star-formation rate agrees with the
true one only at the beginning of the simulation.  Minor
deviations arise later on and are due in part to the production of
high-energy photons in the warm-hot gas, in part to the leakage of
energetic photons out of the galaxy.  The effect of artificially
introduced thin dense shells (with thicknesses smaller than the ChDSs
spatial resolution) is investigated, as well.

\end{abstract}


%

\section{Introduction} 
The main information about physical processes and the physical state
of the interstellar medium (ISM) in star-forming dwarf galaxies (DGs)
are obtained from observed emission line spectra from their nebular
components -- gaseous nebulae or nebulae surrounding the compact
star-forming (SF) region \citep[see the textbooks
  of][]{DS,OsterbrockFerland}.

Line intensity ratios are used to infer the physical state (density,
temperature, chemical composition) of ionized regions.  These
indicators are commonly dubbed diagnostic methods and their use date
back to the seventies \citep{S71,Pagel79}, although research is still
ongoing in this field \citep[see][]{Pilyugin03,PP04,Stas06}.  As it is
well known, \citep[see e.g.][]{OsterbrockFerland} some intensity
ratios (such as [OIII]$\lambda$4363\AA\, versus $\lambda$4959\AA\, and
$\lambda$5007\AA) in range typical for HII regions are much
more sensitive to the electron temperature than to electron
density, whereas others (e.g. [SII]$\lambda$6716\AA/$\lambda$6731\AA\,
and [OII]$\lambda$3729\AA/$\lambda$3726\AA)\, are 
sensitive to the electron density.  In some cases, intensity
ratios are very sensitive to both $T_e$ and $n_e$.  In this
case, two or more diagnostics must be used at the same time to
determine both quantities \citep[see][]{DIAGN,NEBU}.  Physical
conditions inside nebulae can thus be recovered and, from them, the
relative abundances of ions can be determined.

Most of diagnostic methods assume constant electron temperatures and
densities as well as ionic abundances over the whole ionized regime,
although many attempts can be found in the literature to relax this
hypothesis.  The assumption of non-uniformity is necessary for
instance to explain the discrepancies between electron temperatures
found with different diagnostic methods.  In particular, the
temperature fluctuations are characterized by the so-called $t^2$
parameter \citep{Peimbert67}.  A different approach was proposed by
\citet{Mathis98}, who used ratios $f$ between weights of emitting
regions\footnote{$f=(N_2 n_2 V_2)/(N_1 n_1 V_1)$, where $n_1$ and
  $n_2$ are electron densities in emitting regions, $V_1$ and $V_2$
  are their volumes, and $N_1$, $N_2$ are densities of the emitting
  ions} to characterize the temperature inhomogeneities.  Based on
this work, \citet{Stas02} modified the Peimbert's $t^2$ parameter to
allow for temperature inhomogeneities in ionized nebulae.  However, in
all these methods (and in other similar ones), both electron
temperature and density in a {\it single ionization zone} are assumed
to be constant, whereas this assumption is incorrect for most real
nebulae, because density, temperature and chemical abundances are all
inhomogeneously distributed \citep[see e.g.][]{TP05} for the case of
30 Doradus and \citet{Freyer03} for models of wind-blown and
radiation-driven HII bubbles).  Inhomogeneities are much more obvious
if we wish to investigate an entire galaxy, where different gas
phases, characterized by very different temperatures, densities and
metallicities, coexist \citep[see e.g.][]{Walt07,Johnson12}.  It is
thus necessary to perform a more detailed photoionization modelling of
the nebular gas (NG) in DGs that takes into account the spatial
variation of physical properties of the gas.  We can thus obtain
intensity maps of astrophysically relevant emission lines. Using these
maps it is possible to calculate the synthetic spectrum of the
emission lines along different sightlines.

Emission-line diagnostics are also crucial to infer the correct
star-formation rates (SFRs) from galaxies, most commonly from \Ha
emission \citep{Kenn98}.  In fact, Kennicutt's SFR determination is
based on the assumption that photons responsible for the hydrogen
ionization stem from the stellar Lyman continuum (Lyc) alone and
neglects other contributions as e.g.  the cooling radiation of hot
bubbles and the excitation from shock-heated regions.

In this paper we illustrate a way to combine detailed chemo-dynamical
simulations (ChDSs) of DGs with a state-of-the-art multi-component
photoionization model (MPhM), to study the emission properties of
star-forming DGs and compare them with observations.  The idea is to
use the density, temperature and chemical abundances obtained by means
of ChDSs at different locations in the DG and to solve (this is the
task of the MPhM) the ionizing radiation transfer equation as well as
the ionization-recombination, the energy balance, the radiation
diffusion equation and the statistical equilibrium equations in
different parts of the galaxy, using as physical parameters for each
region the results of the ChDS.  An additional crucial ingredient is
the stellar spectrum, which is provided by running a tailored {\it
Starburts99} simulation \citep{SB99}.

The strategy is thus to run conventional ChDSs and to use the MPhM as
a post-process \citep[see also][]{Yajima11,Paar13,Compo13}.  The obvious
advantage of our approach when compared to radiative hydrodynamical
simulations \citep{Yoshida07,Krum12,Hirano14} is that the required
computational time is very short and that not many assumptions need to
be made about radiative transfer.  The serious disadvantage of our
approach is that the radiation field does not feedback on the
hydrodynamics as it should and the energy and pressure in the ChDSs
are purely of thermal origin.

This is the first of a series of papers in which we try to extract as
many useful information as possible from this combined use of
hydrodynamical simulations and photoionization models.  In this first
paper, we will mainly focus on the details of the MPhM and the synergy
between it and the chemo-dynamical simulations.  Moreover, we will
focus on the escape probability of ionizing photons from our model
galaxy, a topic of considerable relevance in the study of the
reionization of the universe, as it is believed that DG-sized objects
at very large redshifts are the main culprits for the reionization of
the universe \citep[see e.g.][]{Vanz12,KC14,Wise14}, although their
contribution is still under debate \citep[see e.g.][]{Fujita03}.
Finally, we will in detail exploit the \Ha emission from galaxies and
the way it has been used in the past to estimate the SFR in galaxies.
As mentioned above, we will revise the estimates of SFRs based on the
Kennicutt's \Ha--SFR relation and we will also offer a possible
explanation for the mismatch of \Ha vs. UV brightnesses \citep{Lee09}
as proposed by \citet{Relano12} for galaxies with SFRs of less than
$\sim$ 0.01 M$_\odot$~yr$^{-1}$.

The paper is organized as follows: in Section 2 the DG simulations are
summarized.  In Section 3 the details of the evolutionary models of
the SF region are calculated; in particular, the Lyc spectrum due to
the DG star formation history is derived.  In Section 4 the MPhM is
thoroughly described and the results of the photoionization models of
different DG angular sectors are analyzed.  In particular, we analyze
the behavior of intensities of several important emission lines
emitted along the different radial directions from the SF region.  In
Section 5 the numerical scheme for the calculation of the synthetic
emission line spectrum (using emissivity maps obtained from MPhM) for
any aperture position will be given.  Moreover, the ability of
different diagnostic methods for the determination of physical
parameters and chemical abundances in the NG is tested.  In the same
Section we consider whether \Ha is a good indicator of the SFR
in galaxies, as it is commonly assumed.  In Section 6 we study in
detail the escape of ionizing photons from our model galaxy.
Moreover, the change in the Lyman continuum spectra due to the
intervening hot gas is studied in detail.  In Section 7 the most
important results of this study are discussed and some conclusions are
drawn.


\section{Description of the chemo-dynamical simulations}
\label{sec:ChDS}
The ChDSs we use in this work are based on the simulations of
\citet{rh13}, hereafter RH13. This work was aimed at simulating dIrrs
with different baryonic masses and initial gaseous configurations, in
order to study the conditions under which a galactic wind could
develop and determine the fate of metals, freshly produced during a SF
episode.  In all RH13 simulations, the galaxy is assumed to be axially
symmetric and rotating about the axis of symmetry, with a frequency
that depends only on the distance from the axis $R$ and not from the
vertical coordinate $z$, in compliance with the so-called
Poincar\'e-Wavre theorem \citep{tass78}.  The numerical code solves
the 2-D gas-dynamical equations in cylindrical coordinates, using a
second-order scheme with flux limiters.  The evolution in space and
time of various chemical elements (H, He, C, N, O, Mg, Si and Fe) is
followed by means of appropriate passive scalar fields.  Detailed
sources of energy and of chemical elements due to supernovae (SNe) of
both Type II and Type Ia and from winds of massive and
intermediate-mass stars are taken into account.  The details of the
numerical scheme and of the implementation of the feedback from dying
stars can be found in \citep{rmd01,recc04,recc06}.

The set of models of RH13, in spite of the fact that they are not
tailored to any specific dwarf galaxy (at variance with the ones of
\citet{recc06} for instance), have a much higher spatial
resolution and a more accurate description of the chemical feedback
than previous simulations of our group.  They are thus ideal to
illustrate the way in which emissivities can be calculated starting
from ChDSs.  We will use, as a reference model, the model L8F of RH13,
i.e. a model with 10$^8$ M$_\odot$ of baryonic mass and a flat initial
configuration.  This model develops a nice and undistorted galactic
wind (see fig. 4 of RH13).  Since one of the aims of our paper is to
study the emissivities from galactic wind regions and escape fractions
of various photons, this model is ideal for our purposes.  We have
also studied the emission properties of the model L8M of RH13.  This
is a model, similar to the reference model L8F, but with a reduced
initial degree of flattening.  Consequently, a galactic wind develops
later in this model compared to model L8F and in a more distorted and
turbulent way (see again fig. 4 of RH13).  In both models, the 
SFR is assumed to be constant during the first 500 Myr of
galactic evolution and equal to 2.67 $\cdot$ 10$^{-2}$ M$_\odot$
yr$^{-1}$.  In spite of the differences, the results obtained with
model L8M are qualitatively similar to the ones obtained with the
model L8F.  They are only more difficult to interpret due to the
enhanced turbulence, therefore we concentrate on L8F.  We do 
expect significant differences with the roundish model L8R though.  
A detailed comparison between emissivities and spectra obtained with 
models L8F, L8M, and L8R will be presented in a forthcoming paper.

\begin{figure}
   \centering
   \includegraphics[width=9cm]{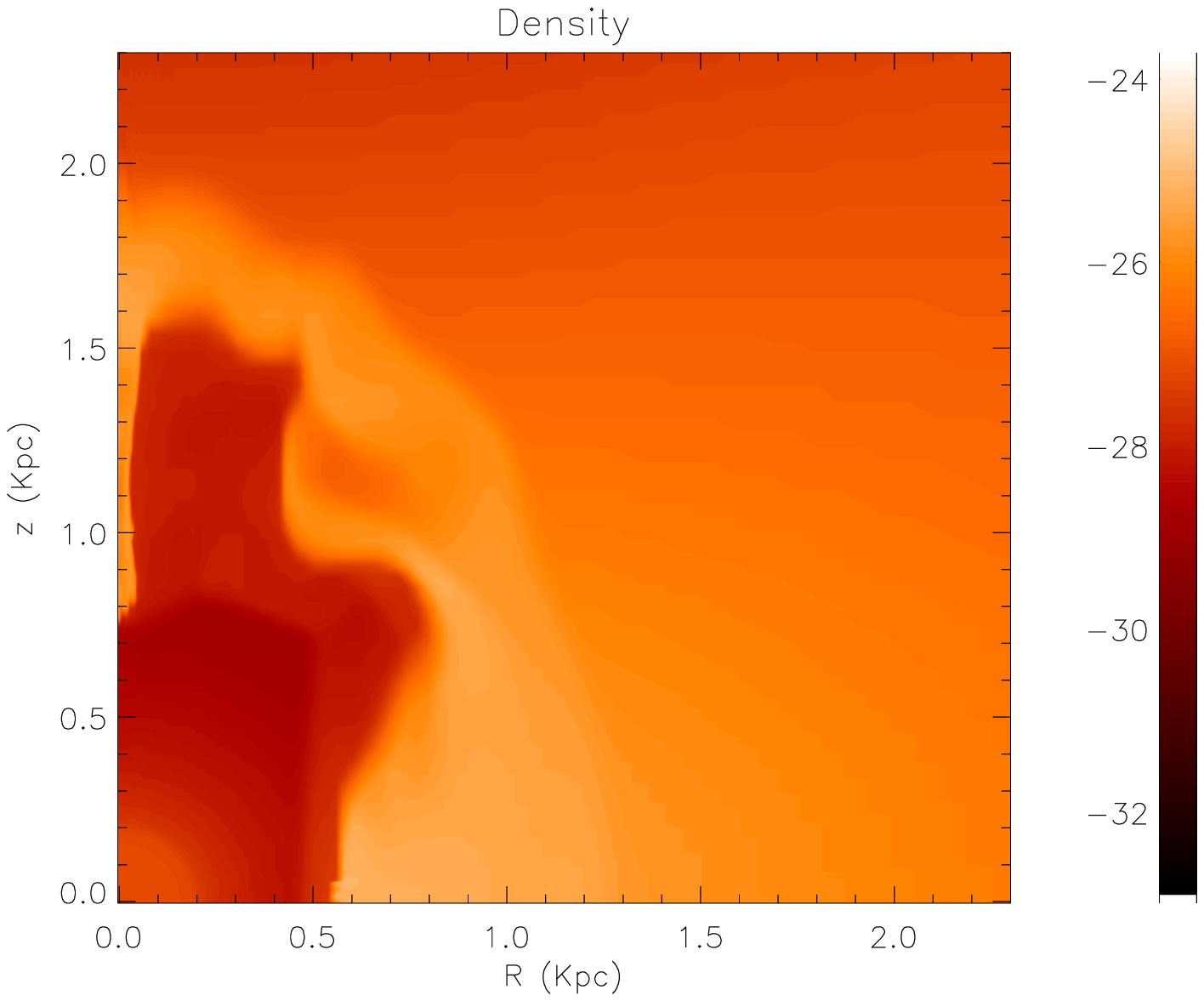}
   \includegraphics[width=9cm]{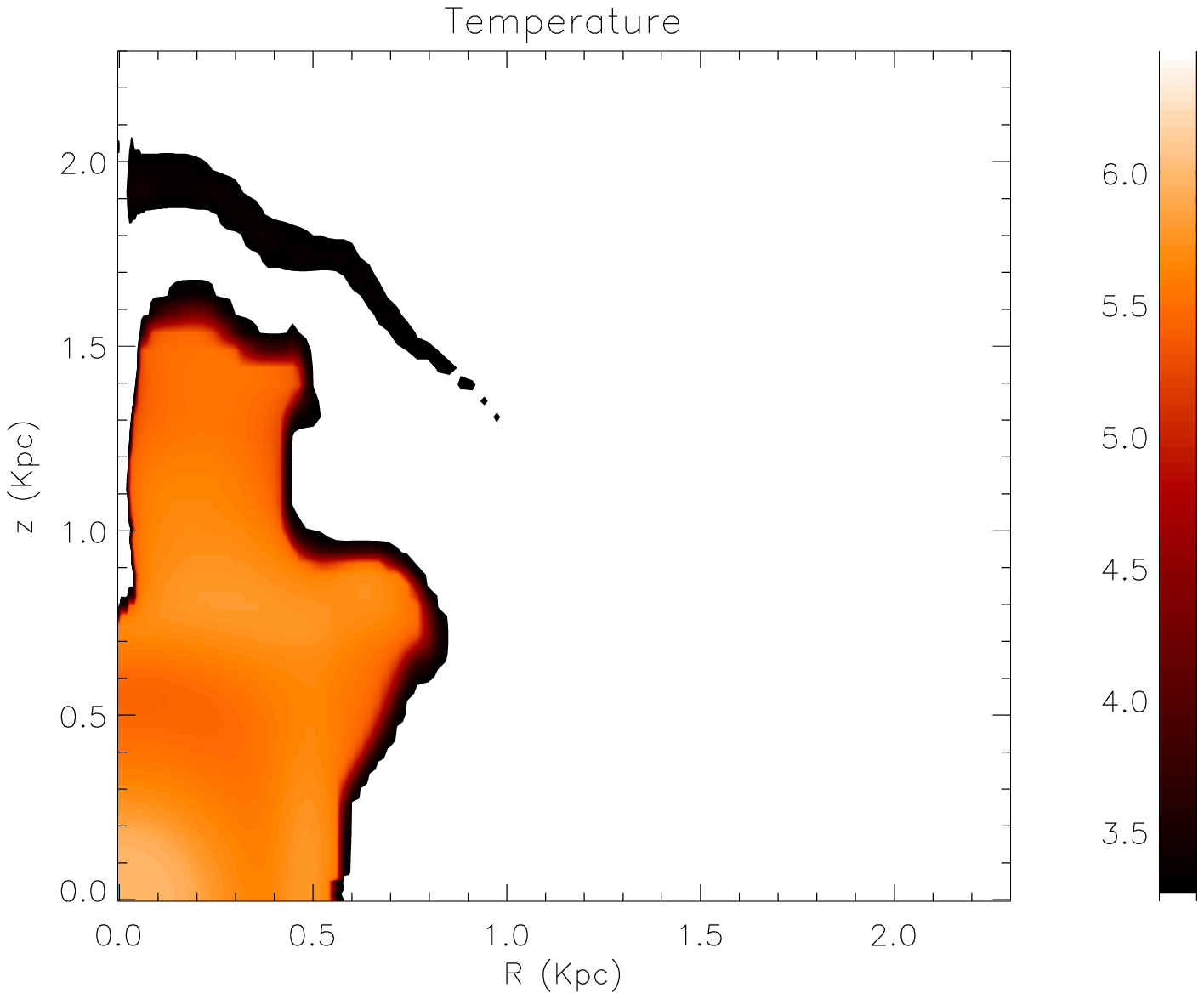}
   \caption{Gas density (upper panel) and temperature (lower panel)
     maps obtained from the L8F model (see Sect. 2 for details on the
     ChDS models), calculated at an evolutionary time of 140 Myr.
     Brighter colors indicate larger gas densities and temperatures.}
   \label{fig:gas_O}
\end{figure}

A typical map of density and temperature resulting from the ChDS of
the model L8F is shown in Figure \ref{fig:gas_O}.

\section{Spectra of the ionizing radiation from stars}
\label{sec:starspec}
The Lyman continuum spectrum originates from the stellar population
and then is modified by the intervening gas.  In order to set the
starting ionizing spectrum, we use tailored Starburst99 (v.7.0.1)
simulations.  The input SFR matches the one adopted in various ChDSs
(see Section \ref{sec:ChDS}). The adopted initial mass function (IMF)
is the Salpeter one.  The Padova AGB evolutionary tracks for Z=0.004
as well as the Maeder wind model were used.  We run the Starburst99
simulations for 200 Myr, with timesteps of 1 Myr.
\begin{figure}
\centering
\includegraphics[width=6.0cm,angle=-90]{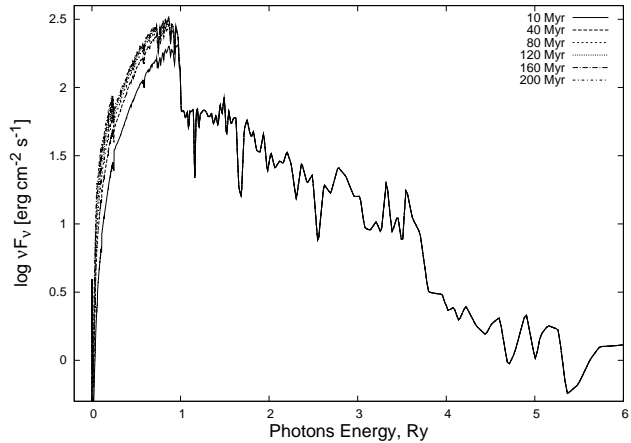}
\caption{Syntethic radiation spectra from the SF region
  at different ages
}
\label{fig:Lyc_SF}
\end{figure}

The total number of the ionizing photons $Q_{ion}$ emitted by the SF
region per unit time weakly depends on the age. The resulting Lyc
spectra at the inner radius of the nebular region are given in
Figure~\ref{fig:Lyc_SF}.

\section{The multi-component photoionization modelling}
\label{sec:MPhM}
In this section we present the algorithm to perform the
multi-component photoionization modelling of NG in DGs.  The results
of such modelling based on ChDSs, such as the emissivity maps for
important emission lines, the transformation of the ionizing
Lyc-spectrum and the predicted emission line spectra obtained along
various radial directions will be considered as well.

\subsection{The MPhM wrapper}

As basis for the photoionization modelling we used G.~Ferland's code 
CLOUDY 08.00 \citep[see][]{Cloudy}.

Looking again at the results of ChDSs (see Fig. \ref{fig:gas_O}), the
first important conclusion we can draw is that the ionizing gas can be
divided into two main parts.  The first one is the cavity carved by
supernova explosions and stellar winds around the SF region (superwind
region -- SWR).  This component contains very rarefied and hot gas.
Also the abundance of heavy elements is very high.  Since the MPhM
does not treat shock-heated gas, the temperature in this region can
only be determined by the ChDS.  The second component is the
supershell swept up by the propagating shock (hereafter called
'wall').  It is characterized by much higher gas densities, as well as
by heavy-element abundances typical for the galactic ISM.  The 'wall'
corresponds thus to the outer edge of the SWR.  It is important to
point out that, in spite of the larger densities and the relatively
low temperatures, a significant fraction of the wall is ionized
because the leakage of ionizing photons from the inner region of the
galaxy is significant (see below).  Beyond the wall, the gas is
unperturbed and it is heated mainly by photoionization.  Therefore,
the temperature in this component can be obtained as a solution of the
photoionization thermal energy-balance equation.

The second important conclusion from the analysis of ChDS results is
that the maps of density and temperature are very complex.  In spite
of the axial symmetry, the whole galaxy can not be approximated by a
simple geometrical figure.  Therefore, it is impossible to use any of
the geometrical shapes offered as options in CLOUDY.  Obviously, a
more complex strategy to simulate the gas emission is required, in
order to take into account the very large variations of density,
temperature and chemical composition across the galaxy 
(see also the Introduction for alternative strategies).

\citet{Morisset} (see also references therein) developed the so-called
pseudo-3D extention {\it Cloudy\_3D} of CLOUDY, in order to model the
emission of nebular objects in 3D geometry, using many 1D
photoionization models.  This is an appropriate strategy to model
objects like diffuse ionized gas surrounding star-formation regions.
Nevertheless, we decided not to use this code and to write our own
specialised driver MPhM (multicomponent photoionization modelling) to
calculate the multicomponent photoionization models for the following
reasons: 1) We are not using the standard CLOUDY release but a
customized version.  In particular, it was necessary to update the
core of CLOUDY to distinguish between SWR and ordinary nebular gas.
2) To generate brightness maps end emission line profiles,
\citet{Morisset} developed IDL-procedures, but IDL is not a free
software and therefore it is not available to all astrophysicists who
are interested in such models.  Therefore, we developed our own code
{\it DiffRay} (described below) to calculate the emission line spectra
in synthetic apertures with user-defined sizes and positions.  We plan
to make the MPhM driver as well as the code {\it DiffRay} freely
available to all the astrophysical community.  3) We are continuously
upgrading MPhM (and also the core of CLOUDY) to better adapt it to the
study of the nebular emission in galaxies.

\begin{figure}
   \centering
   \includegraphics[width=9cm]{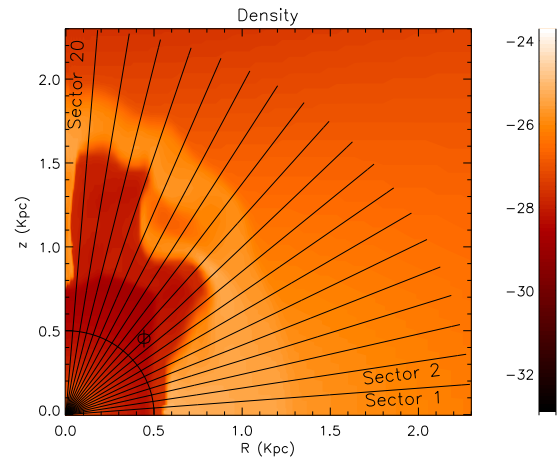}
   \caption{Scheme of the division of the computational box into
     sectors and components.  Sectors are numbered from 1 to 20, with
     Sector 1 being the one adjacent to the galactic plane (i.e.
     adjacent to the R-axis) and Sector 20 the one adjacent to the
     axis of symmetry (i.e. adjacent to the z-axis).}
         \label{fig:ModSketch}
\end{figure}

In the MPhM approach, the whole computational box of the ChDS is
divided into angle sectors, namely solid angles drawn from the origin
of coordinates, which is the location of the central star cluster (see
Figure \ref{fig:ModSketch}).  As already mentioned, ChDSs are
calculated in 2D geometry assuming cylindrical symmetry.  This implies
in particular that only one quadrant in the $R-z$ plane must be
calculated.  We divide this quadrant into 20 angle sectors (see Figure
\ref{fig:ModSketch}).  In what follows, the sectors will be numbered
from 1 to 20, with Sector 1 being the one adjacent to the galactic
plane and Sector 20 the one adjacent to the z-axis (the axis of
symmetry).  Each sector is in turn uniformly divided into 200 radial
components.  The radial extent of each component is 11.5 pc.  The gas
at radial distances larger than 2.3 kpc has been considered as part of
the intergalactic medium.

A simple coordinate transformation and an interpolation allow us then
to map the ChDS results into the MPhM frame.  In particular, in each
of the sectors and components, the following input data have been
derived and stored
\begin{itemize}
\item  the hydrogen density;
\item the electron temperature (only in the SWR, i.e. in the innermost part of
  sectors);
\item the chemical abundances of 9 chemical elements (H, He, C, N, O,
  Mg, S, Si, Fe).
\end{itemize}

It is worth stressing that CLOUDY itself divides the computational
volume into zones which are usually smaller than our above defined
components.  Therefore, CLOUDY performs a further interpolation on the
input data.

The edge between hot SWR and wall is defined by the locus of points
across each sector where the density gradient is highest. In most of
the wall (and also in external regions), the temperature derived from
the ChDS is below $\sim$ 4000 K.  For these components, we rely on the
temperature derived by CLOUDY, as the heating effect due to radiation
can be very significant.  A detailed analysis of such situation is
presented in the following subsection.

In our models, two kinds of ionizing radiation are taken into account.
The first is the stellar emission from the SF region.  In
Fig.~\ref{fig:Lyc_SF} the synthetic ionizing radiation spectra are
shown, depending on time.  The second kind of ionizing radiation is
the diffuse one appearing in the NG, mainly due to the radiative
recombination onto the ground level of the ions H$^+$, He$^+$,
He$^{++}$, onto the second level of He$^{++}$ as well as by $Ly
\alpha$ transitions of HeI and HeII.  The diffuse ionizing radiation
is calculated using the "outward-only" approximation, which is the
default in CLOUDY.  According to this approximation, the diffuse
radiation flux from each component, (taking into account all important
sources of opacity), is added to the incident continuum flux from the
central SF region.  The radiation from each component is allowed to
propagate only outwards.  Thus, each angular sector in the present
MPhM approach is "isolated" from the ionizing radiation from the
others.  This is clearly a simplification, made in order to reduce the
computational costs.  More advanced solutions are currently being
developed and tested.

We perform MPhM calculations in the time interval from 10 to 200 Myr
with timesteps of 10 Myr.  At each time step, we take a different ChDS
output and a different stellar spectrum, in order to keep pace of the
changes occurred in the gaseous and stellar component of the galaxy.

\subsection{Photoionization modelling of sectors}
\label{subs:sectors}

\begin{figure*}
\centering
\includegraphics[width=5.8cm,angle=-90]{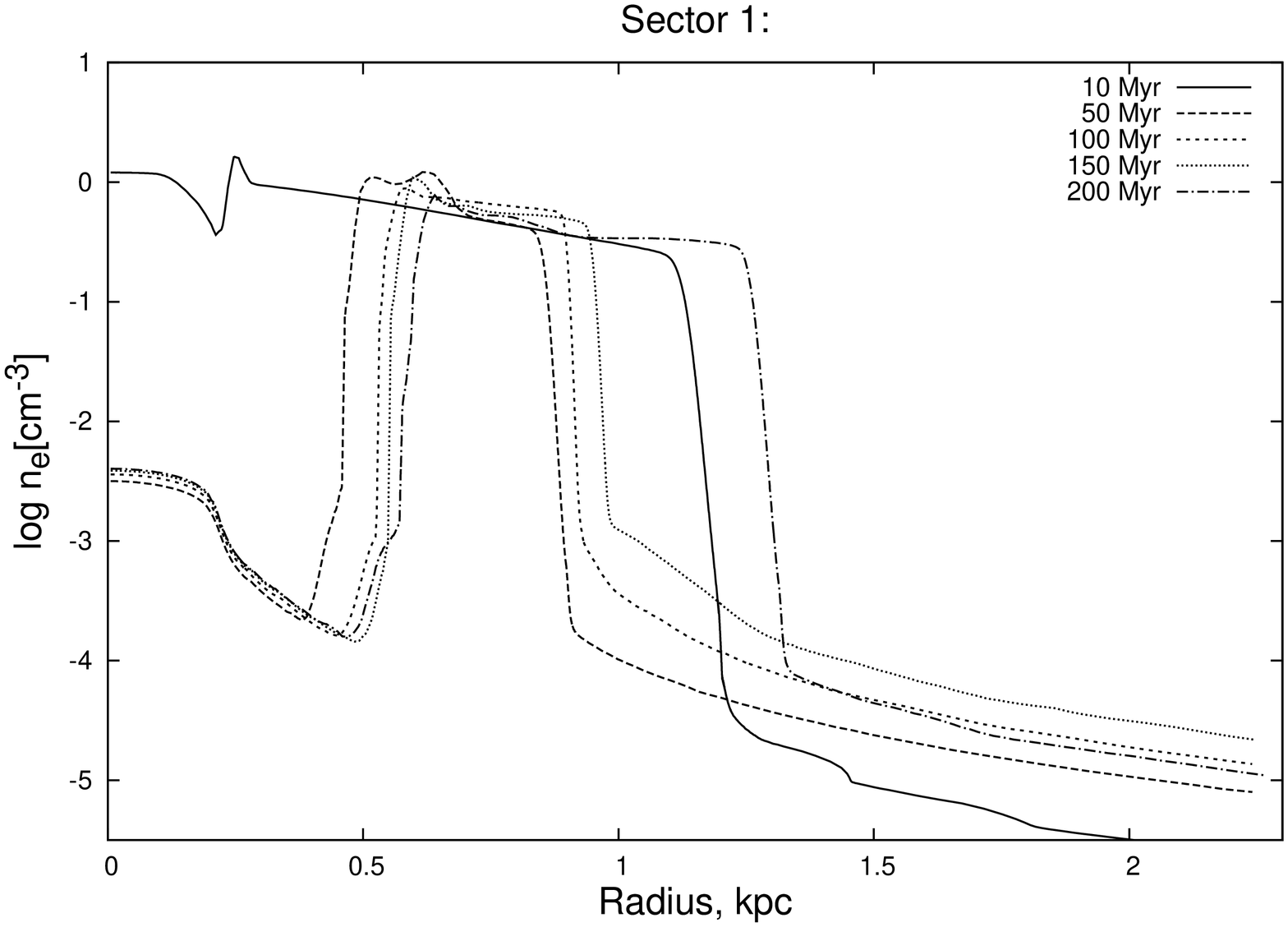}
\includegraphics[width=5.8cm,angle=-90]{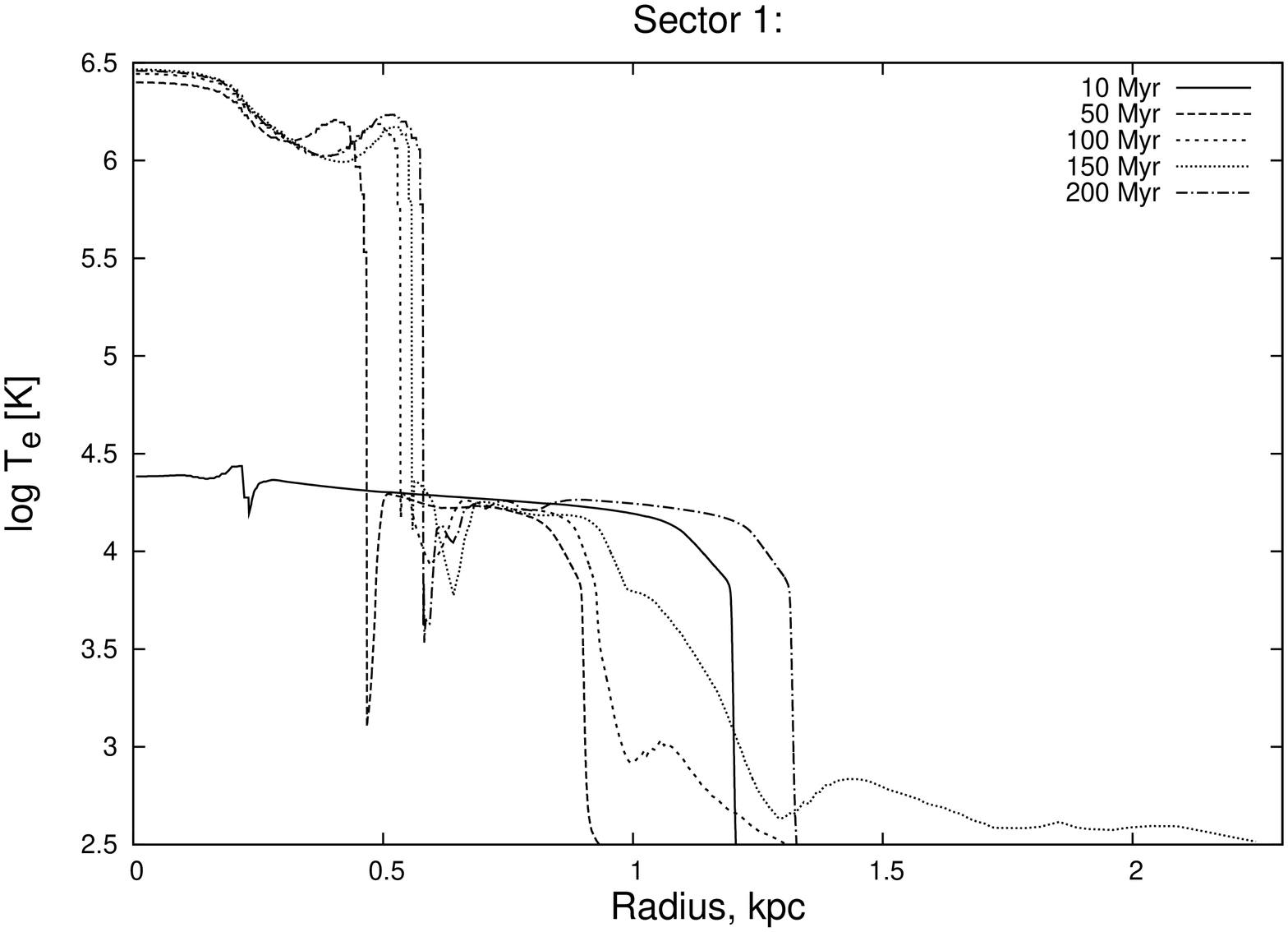}\\
\includegraphics[width=5.8cm,angle=-90]{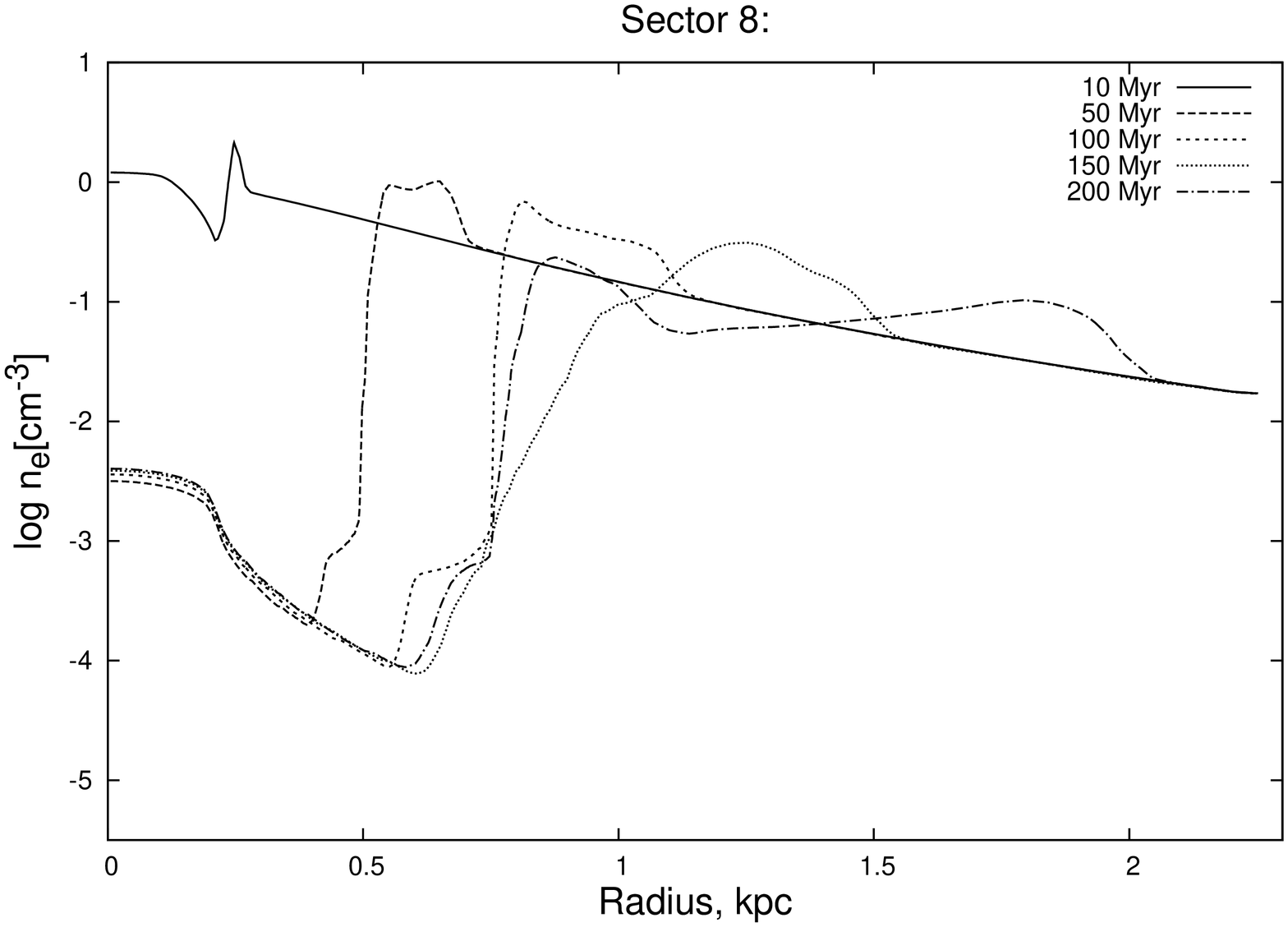}
\includegraphics[width=5.8cm,angle=-90]{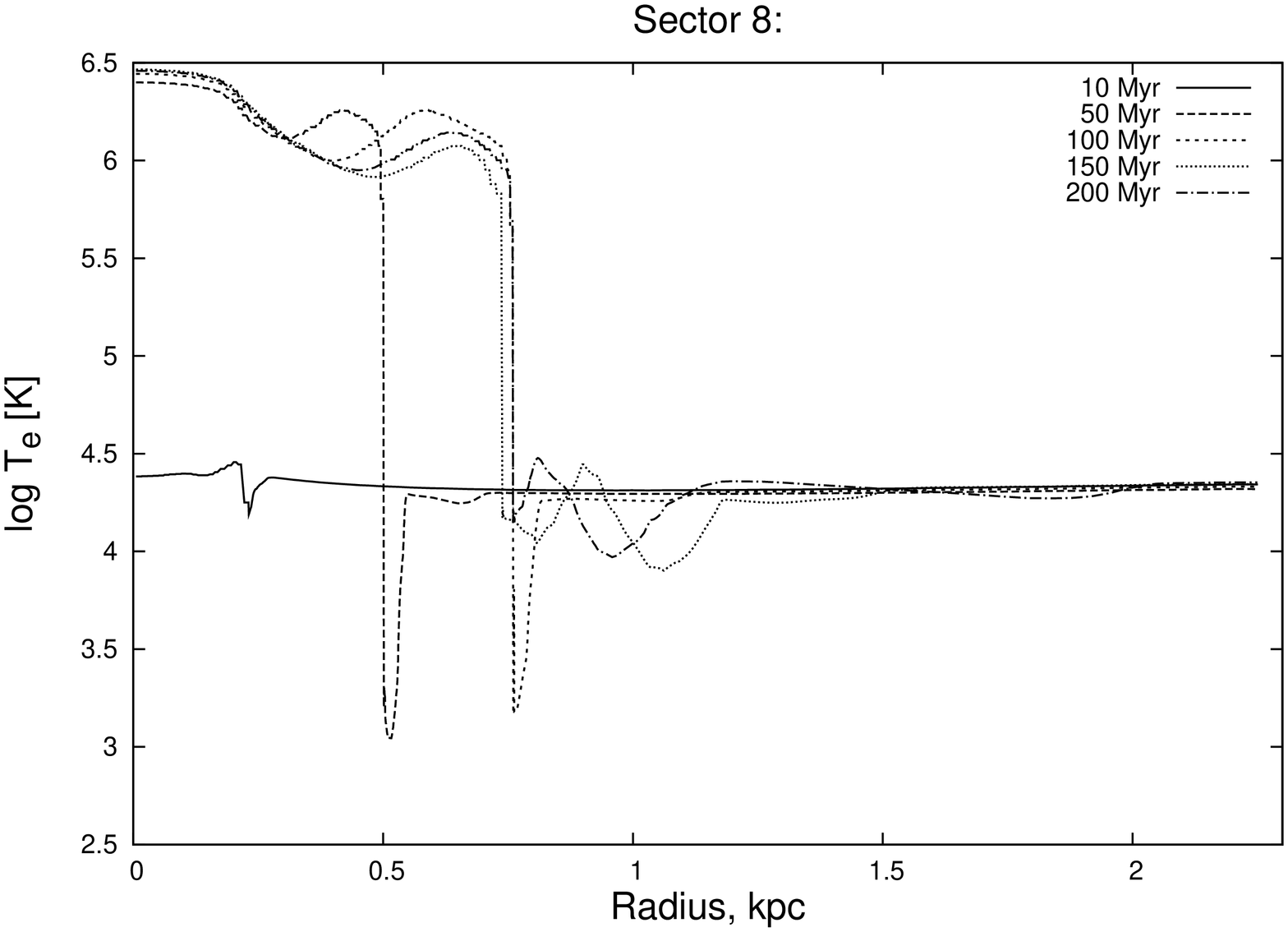}\\
\includegraphics[width=5.8cm,angle=-90]{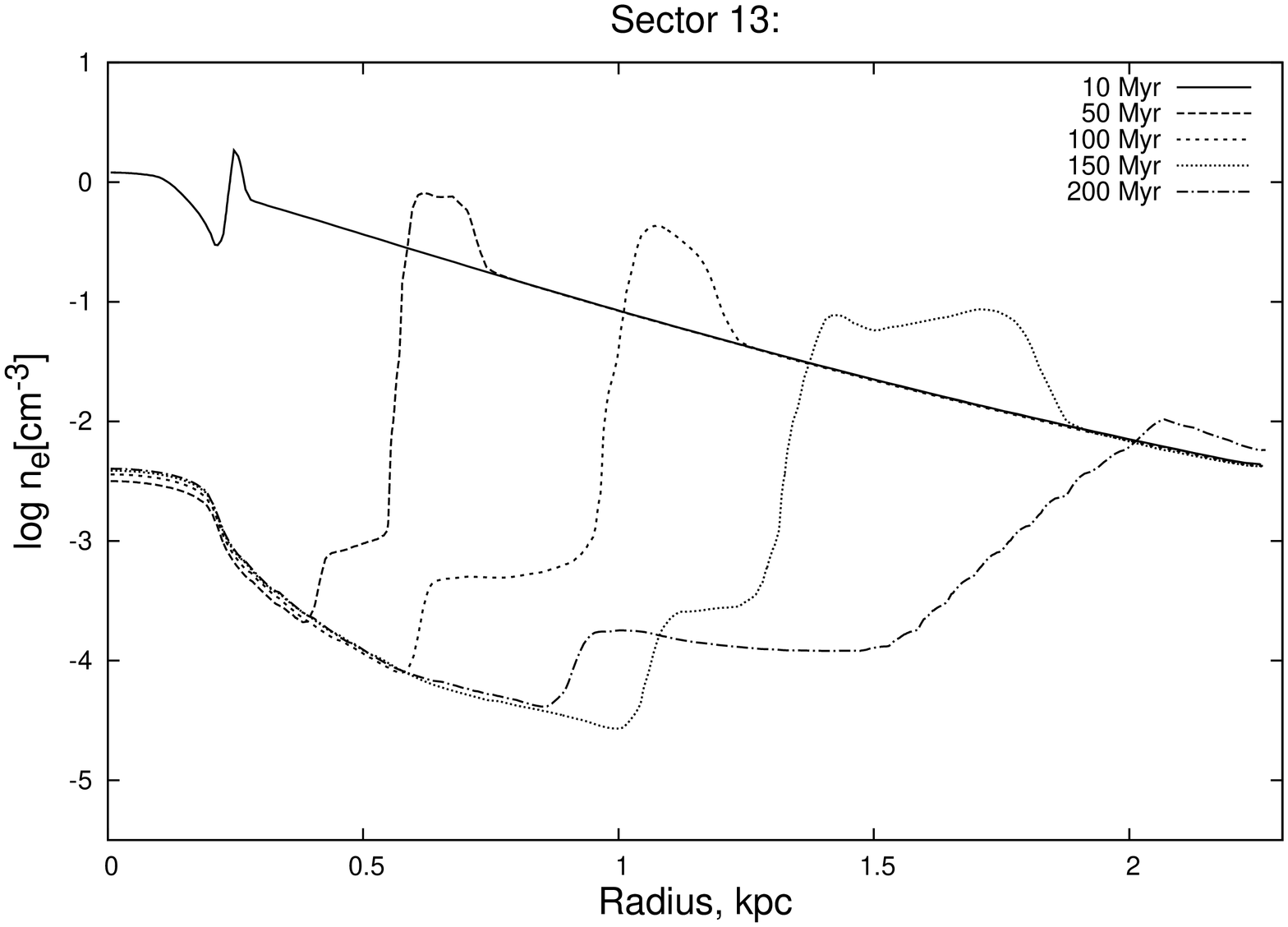}
\includegraphics[width=5.8cm,angle=-90]{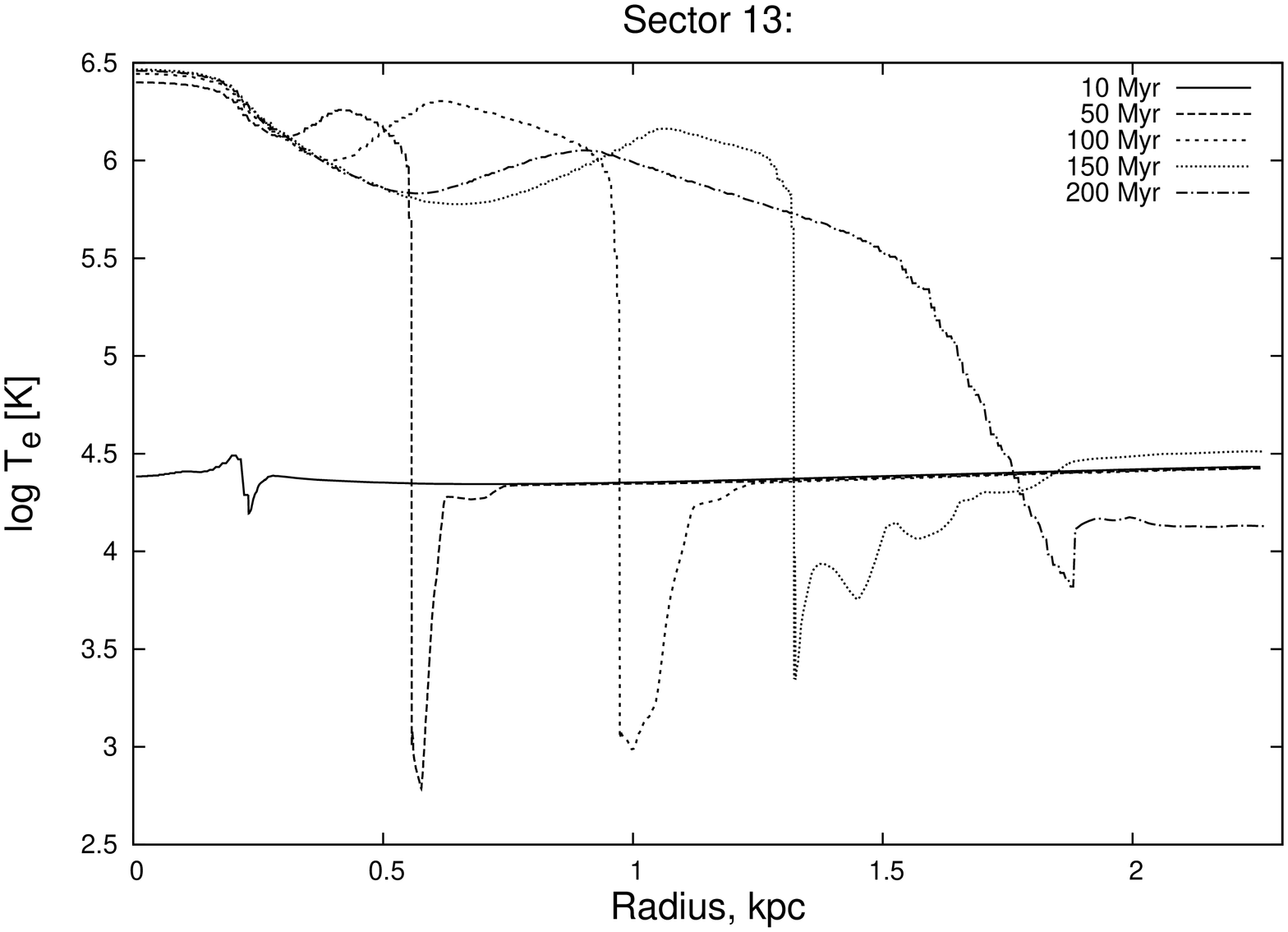}\\
\includegraphics[width=5.8cm,angle=-90]{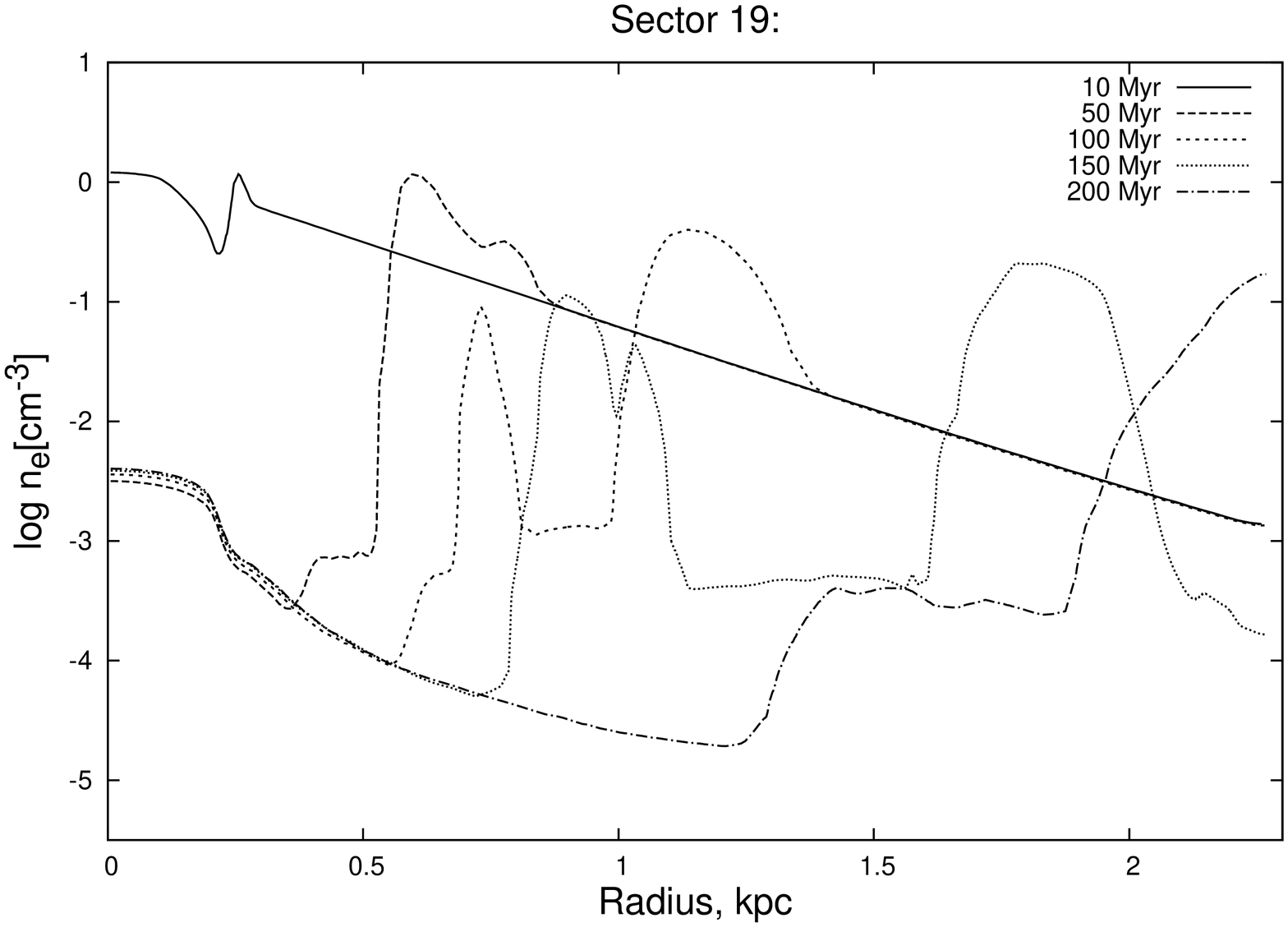}
\includegraphics[width=5.8cm,angle=-90]{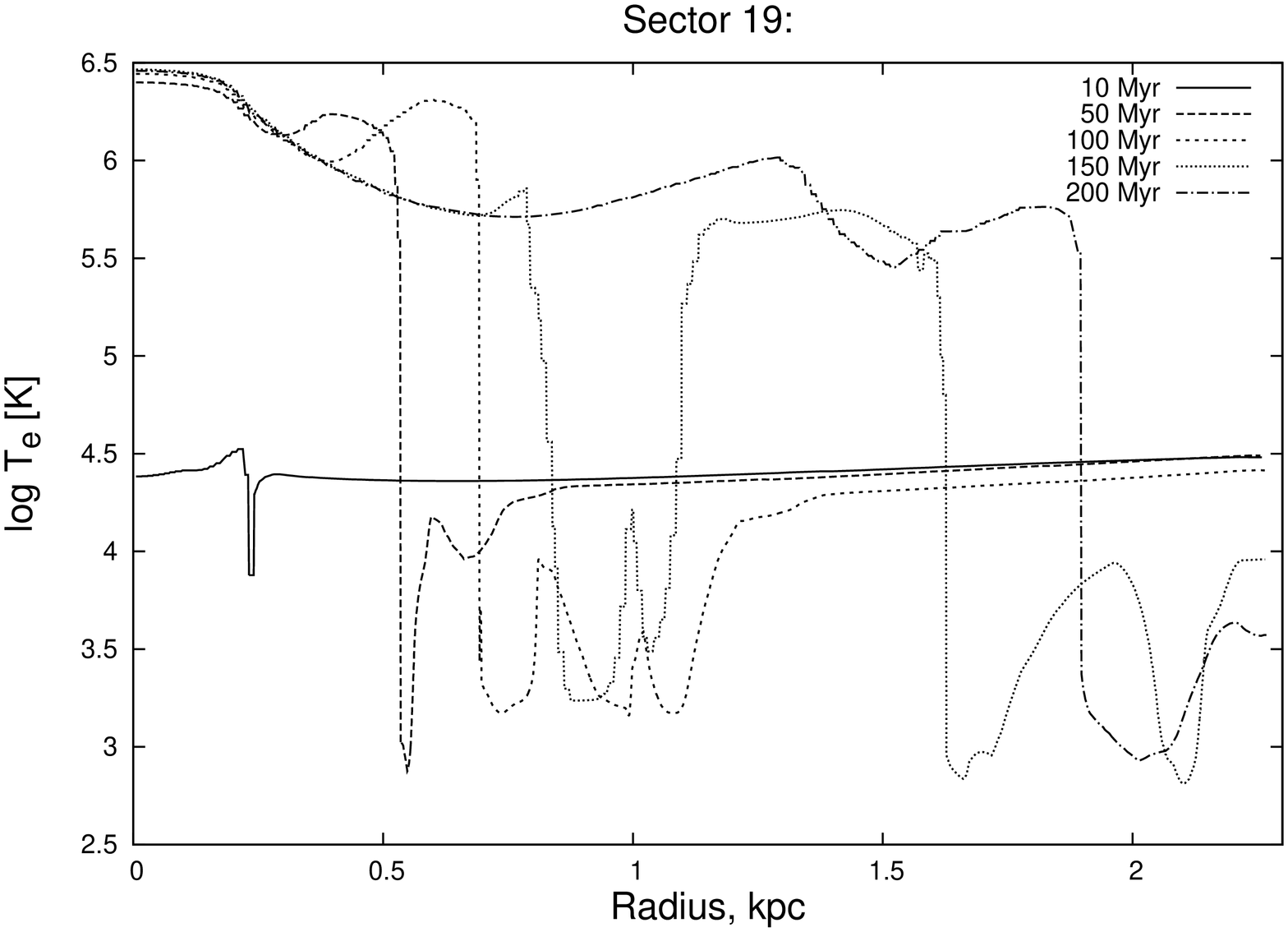}
\caption{Variation of the electron density and temperature across 
    some angular sectors, as a function of radial distance from the
  center.  Different lines indicate different evolutionary times.  The
  calculation refers to the ChDS model L8F (see Sect. \ref{sec:ChDS}
  for details).}
\label{fig:NeTe}
\end{figure*}

As mentioned in Section \ref{sec:ChDS}, we will mostly focus on MPhM
calculations based on the ChDS L8F.  In Fig. \ref{fig:NeTe} the
variation of the electron density and temperature across some
angular sectors are drawn, as a function of radial distance from the
center.  Different lines indicate different evolutionary times.  As a
reminder, Sector 1 runs across the galactic plane, therefore the hole
of low density and high temperature (the SWR in our nomenclature) is
less extended. Conversely, Sector 19 lies close to the
axis of symmetry, i.e. along the direction of steepest pressure
gradient.  Along this direction a galactic wind develops faster and
thus the SWR is more extended (see also Fig.  \ref{fig:gas_O}).

In Fig. \ref{fig:NeTe} one can clearly see the edge between the SWR,
characterized by high temperatures and low densities, and regions of
gas still at lower temperatures and at higher densities.  In some
cases however some oscillations in the temperature profiles can be
discerned, with colder regions preceding hotter ones.  This is due to
the turbulent nature of the galactic outflows and to the adiabatic
cooling of the outflowing gas.  This can be clearly seen in Fig.
\ref{fig:gas_O}.  A ridge of shock-heated gas stretches from (R,z)
$\simeq$ (0, 2) kpc to $\simeq$ (0.8, 1.5) kpc.  This gas has a
temperature higher than the background one.  Behind it, there is a
strip of adiabatically cooled gas.  Moreover, a turbulence-generated
eddy starts developing at (R,z) $\simeq$ (0.5, 1.4) kpc.  It is thus
possible that some sectors, at some evolutionary times, intercept
regions at different temperatures, generating thus the temperature
oscillations visible in Fig. \ref{fig:NeTe}.  This effect can be
particularly significant from Sector 10 onwards.  Because of these
inhomogeneities in the temperature distribution, we allow CLOUDY to
calculate the electron temperature any time the temperature derived
from the ChDS drops below 4000 K.  It can also be seen that for sector
1, containing the densest and thickest 'wall' element, the outer
ionization front occurs at distances from 0.8 to 1.3 kpc, depending on
age. This is the outer ionization front due to the absorption of
ionizing photons in the gas.  This means that in radial directions of
low-number sectors the gas opacity is large and a very small fraction
of the ionizing radiation propagates beyond the wall.

The MPhM allows to calculate the emissivity maps for important nebular
emission lines, that can be used to calculate the corresponding
radiative fluxes along any direction.  This is necessary in order to
obtain theoretical spectra to compare with observed ones.

\begin{figure*}
\centering
\includegraphics[width=5.9cm,angle=-90]{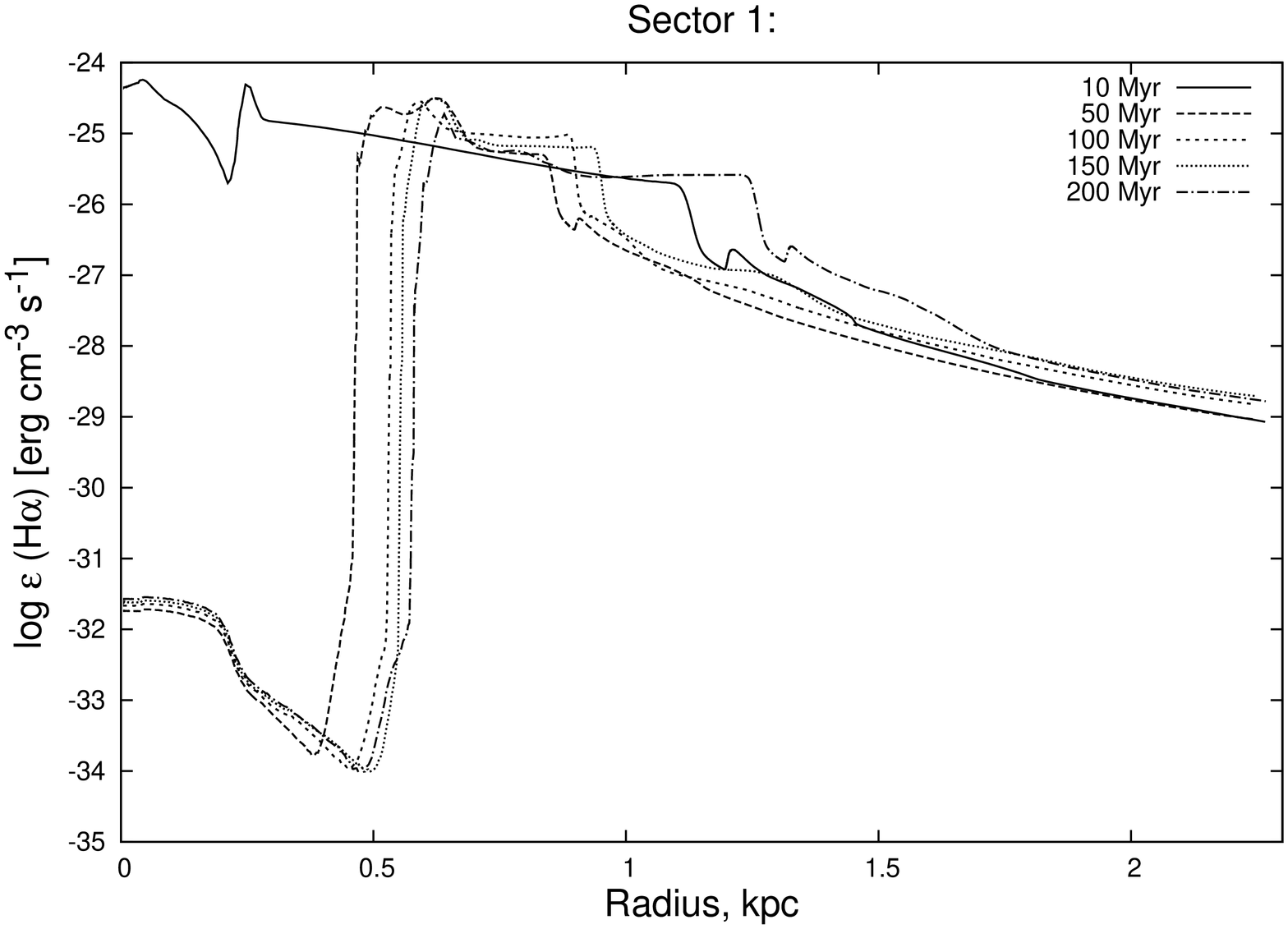}
\includegraphics[width=5.9cm,angle=-90]{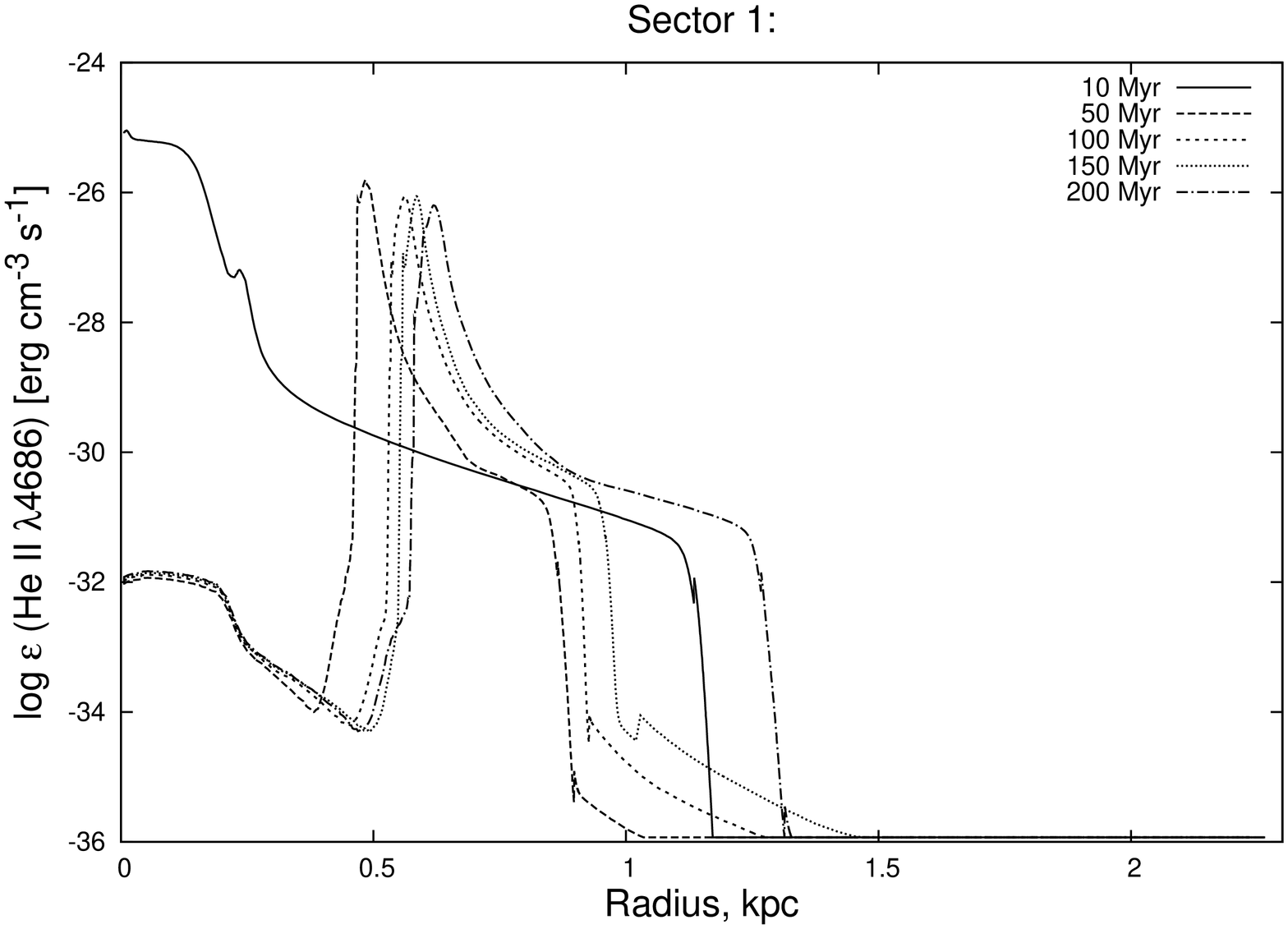}\\
\includegraphics[width=5.9cm,angle=-90]{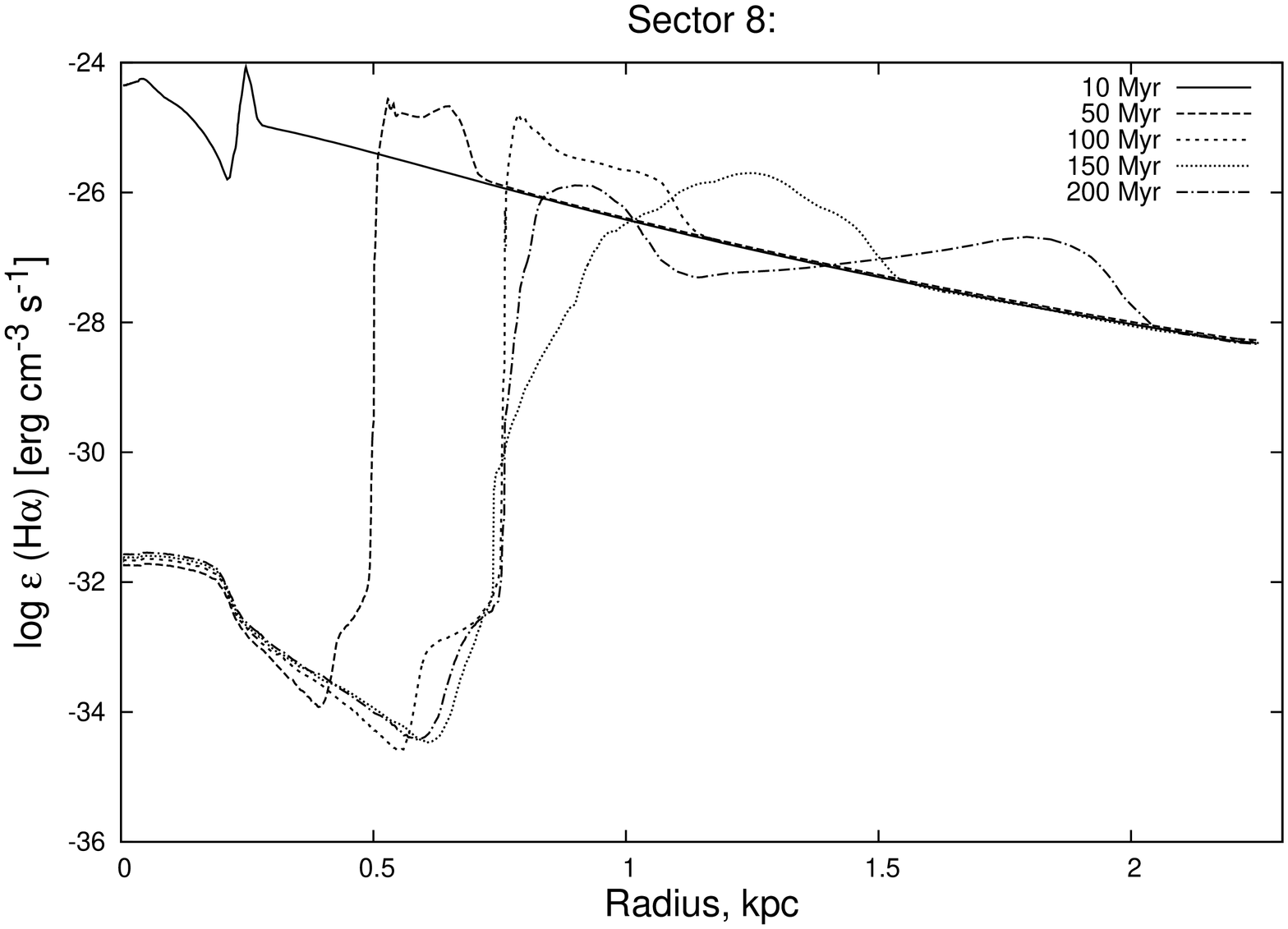}
\includegraphics[width=5.9cm,angle=-90]{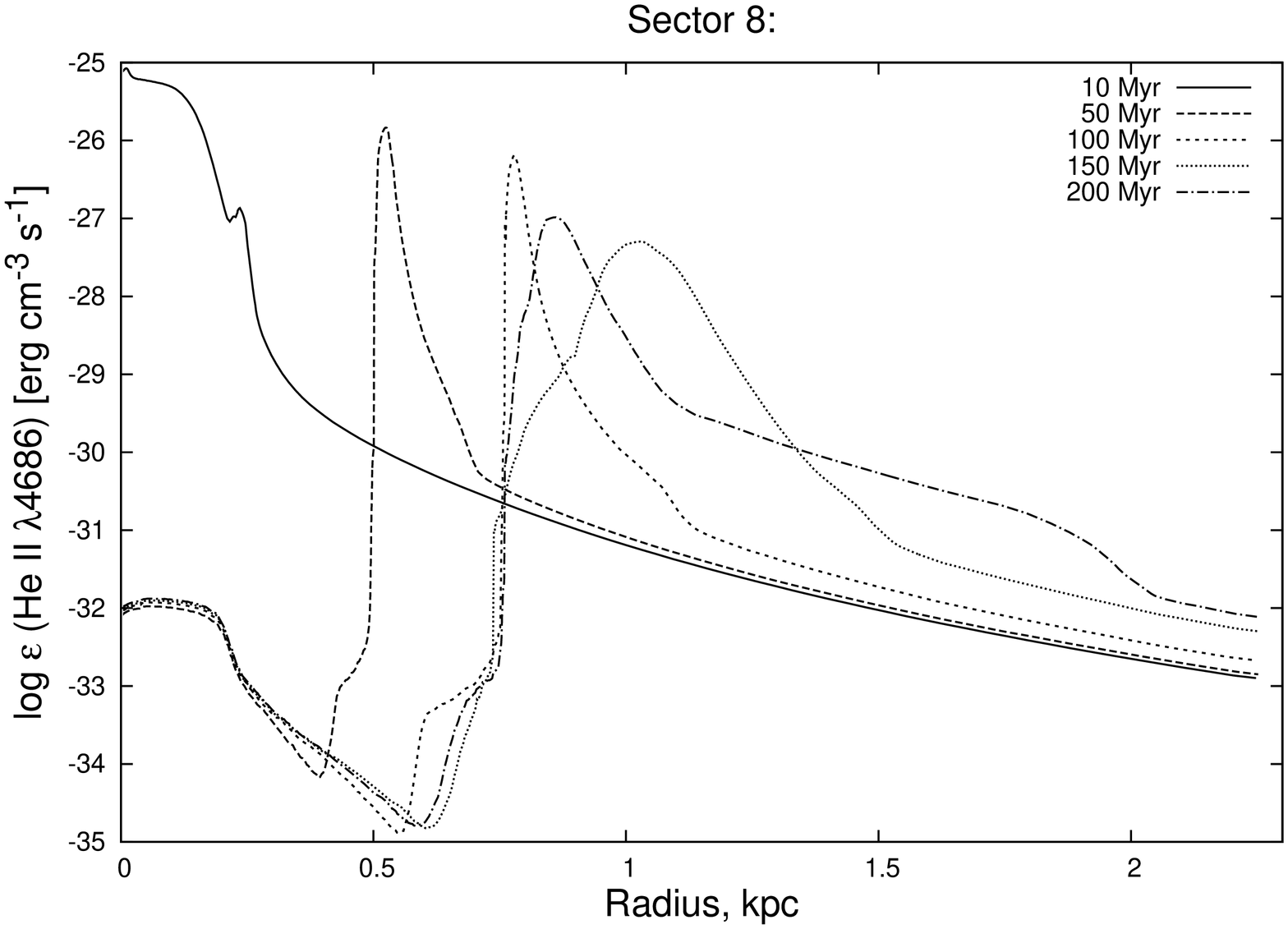}\\
\includegraphics[width=5.9cm,angle=-90]{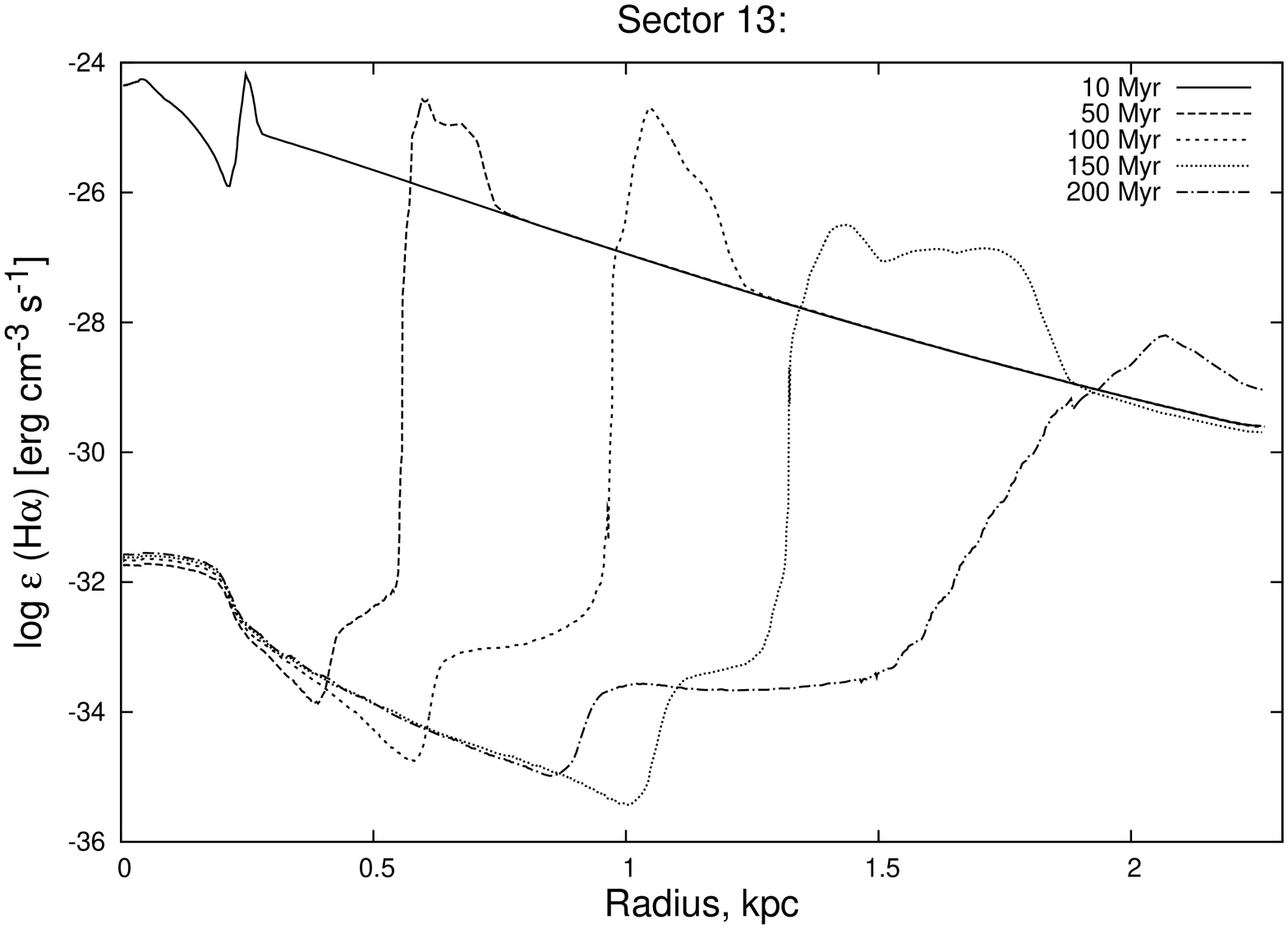}
\includegraphics[width=5.9cm,angle=-90]{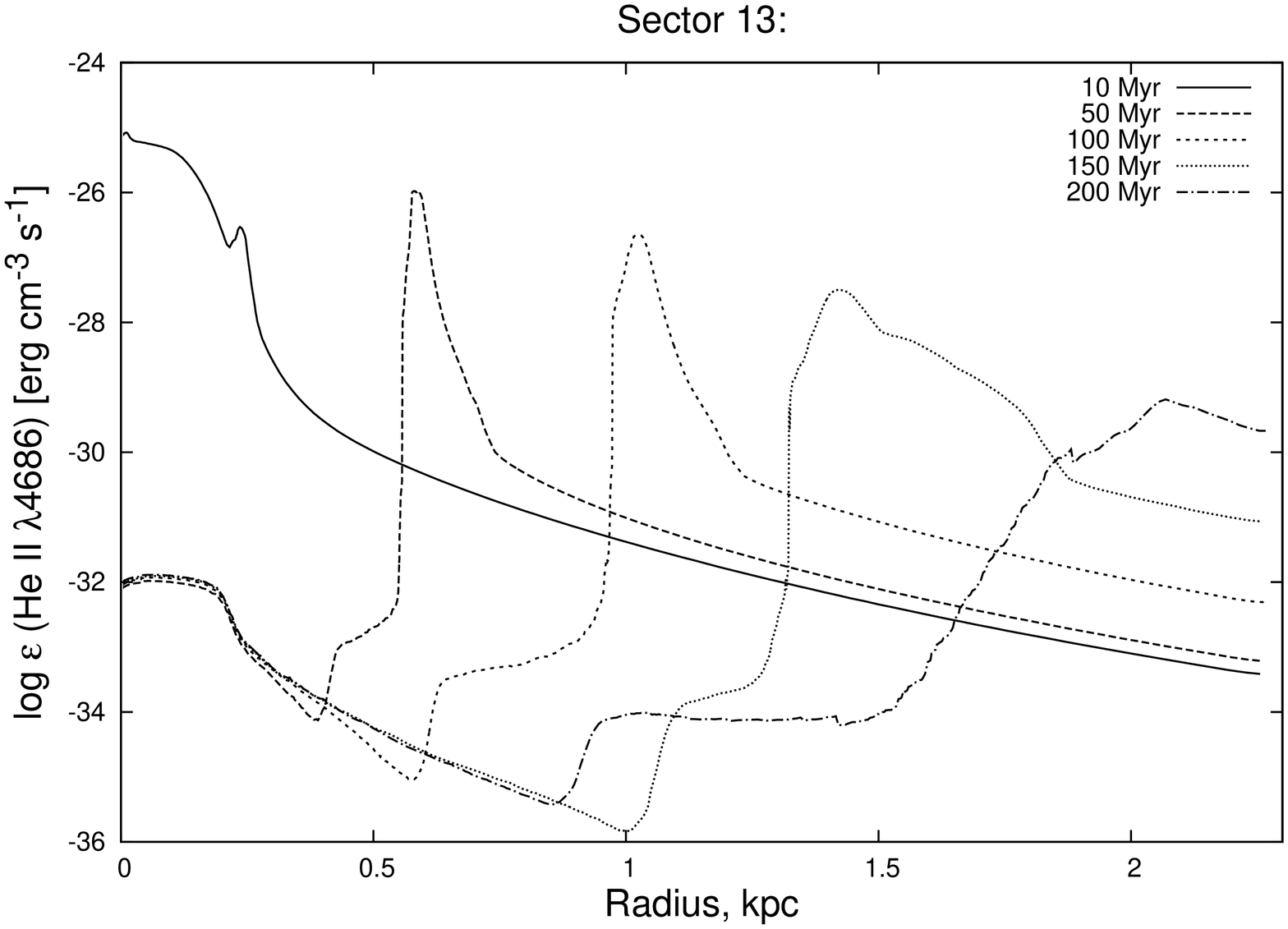}\\
\includegraphics[width=5.9cm,angle=-90]{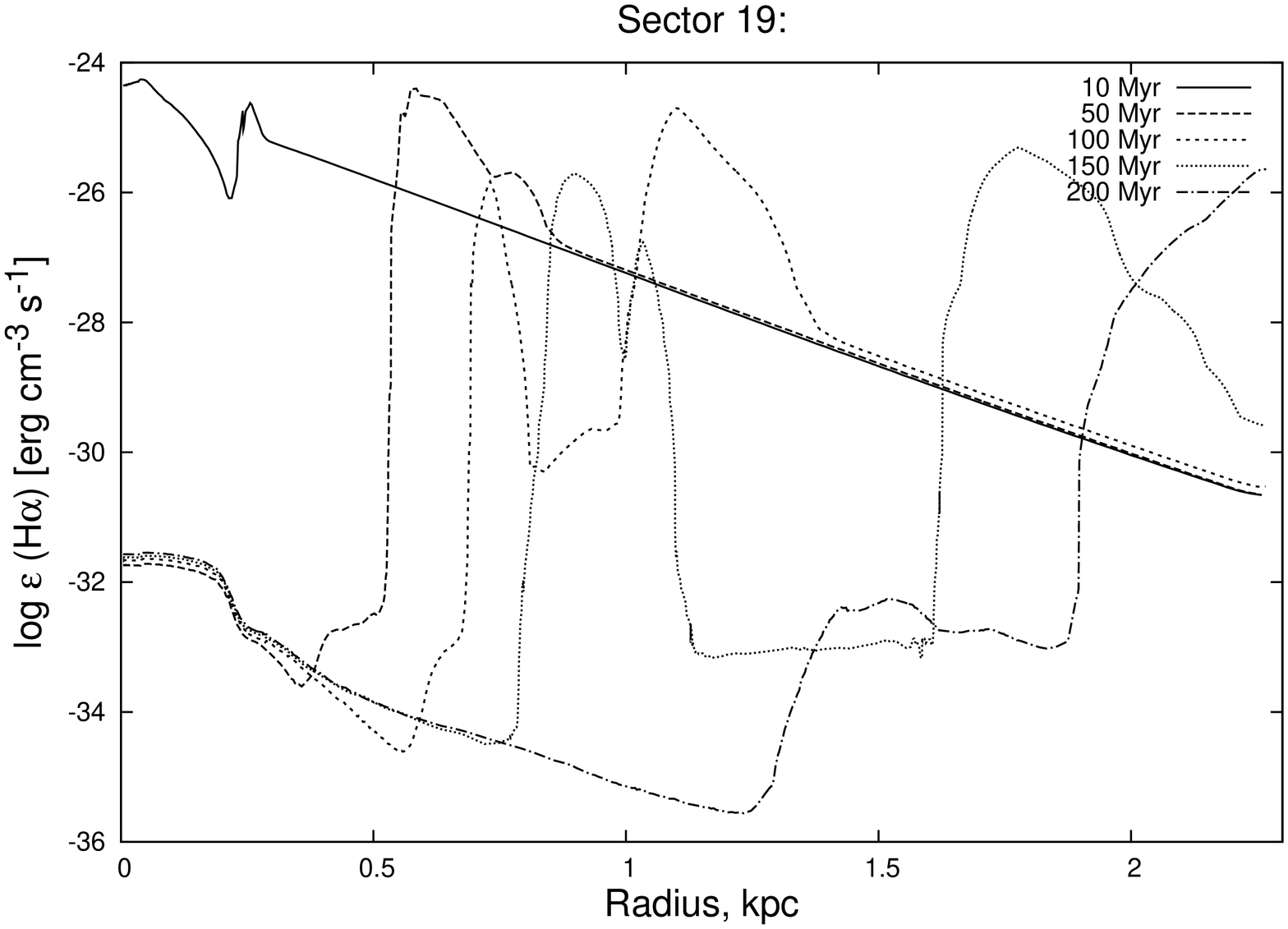}
\includegraphics[width=5.9cm,angle=-90]{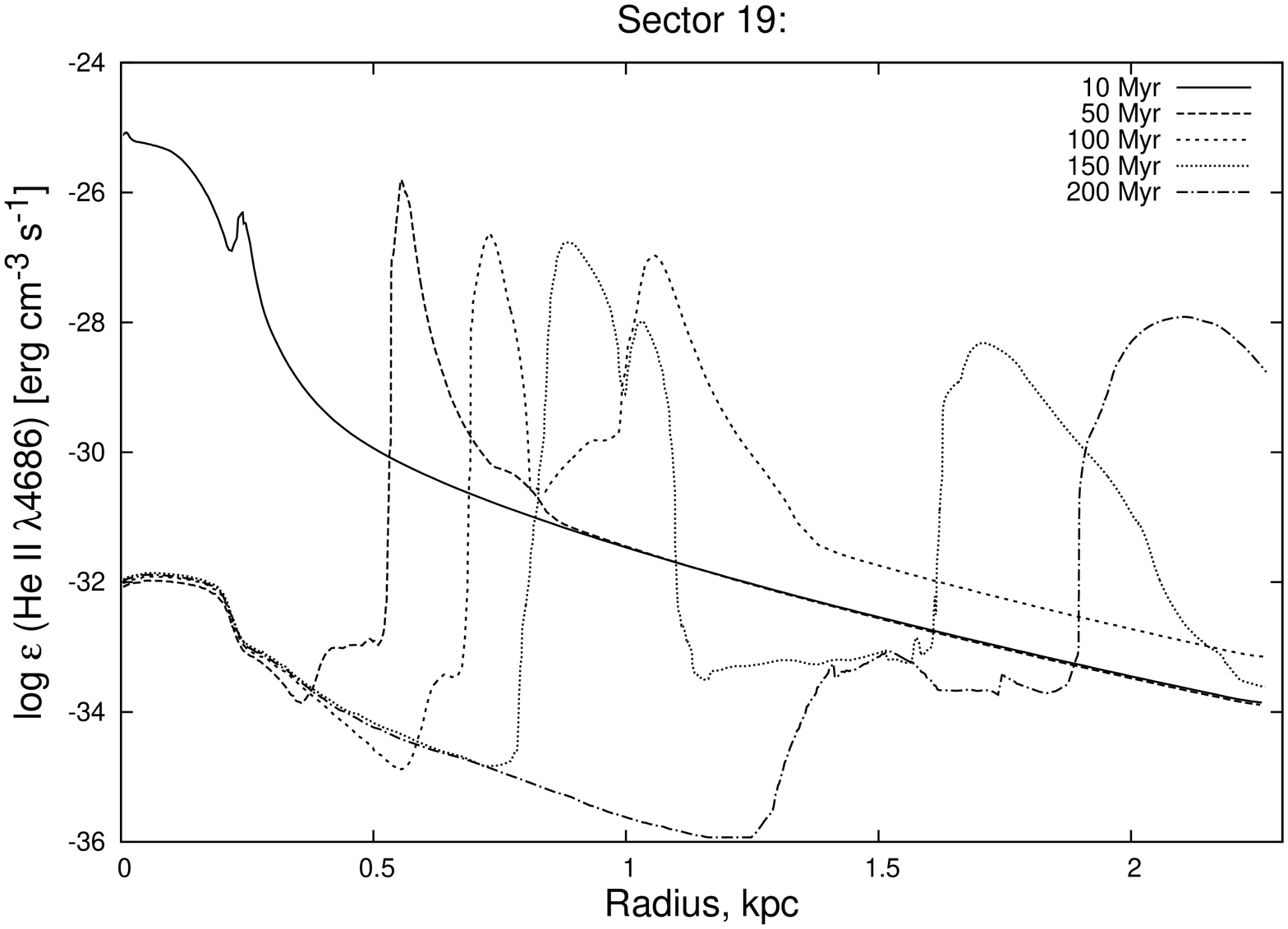}\\
\caption{Radial variation of the emissivities of the recombination
lines \Ha and $\lambda$ 4686 HeII for selected sectors and ages.}
\label{fig:HaHeII}
\end{figure*}

\begin{figure*}
\centering
\includegraphics[width=6cm,angle=-90]{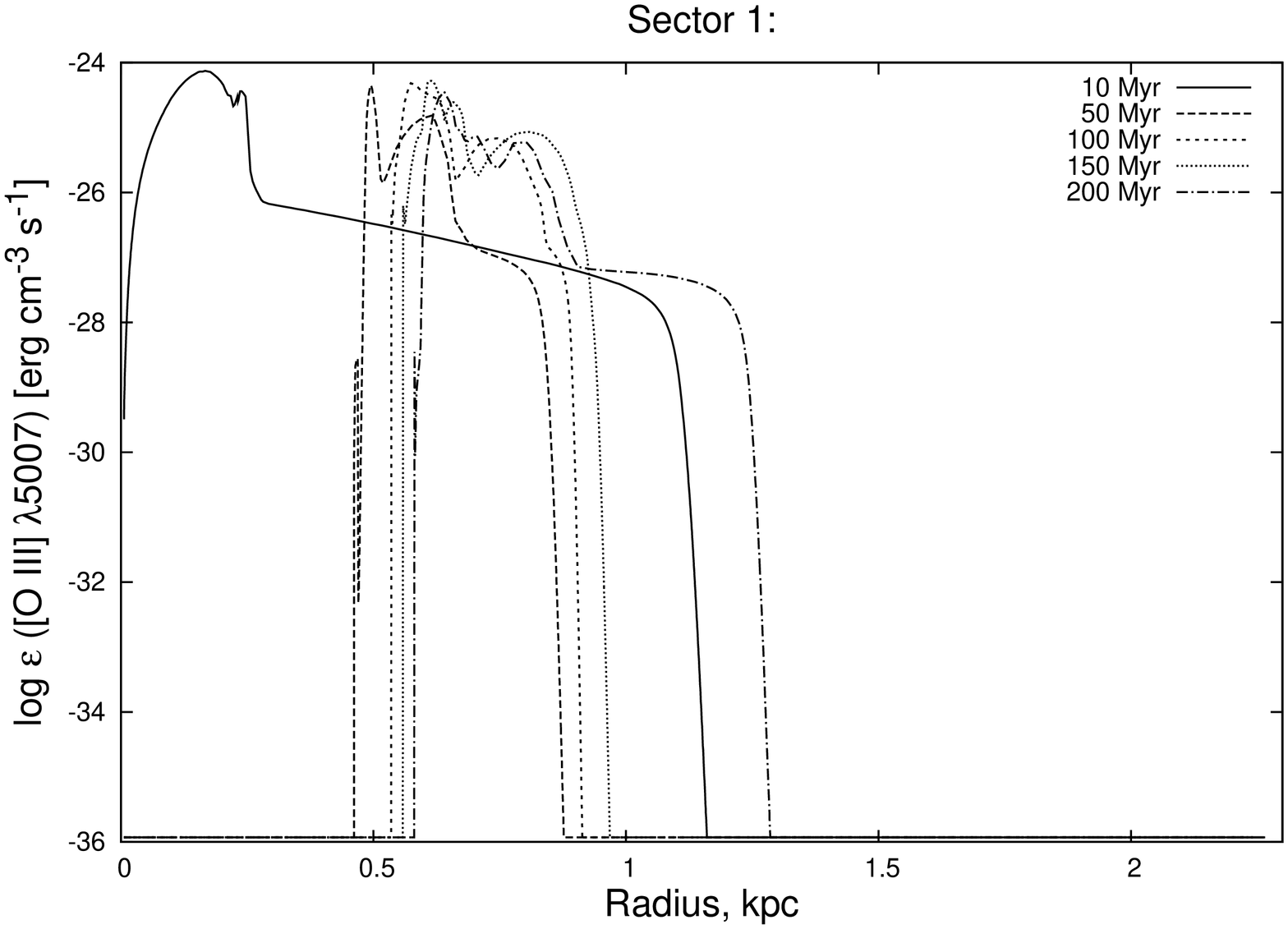}
\includegraphics[width=6cm,angle=-90]{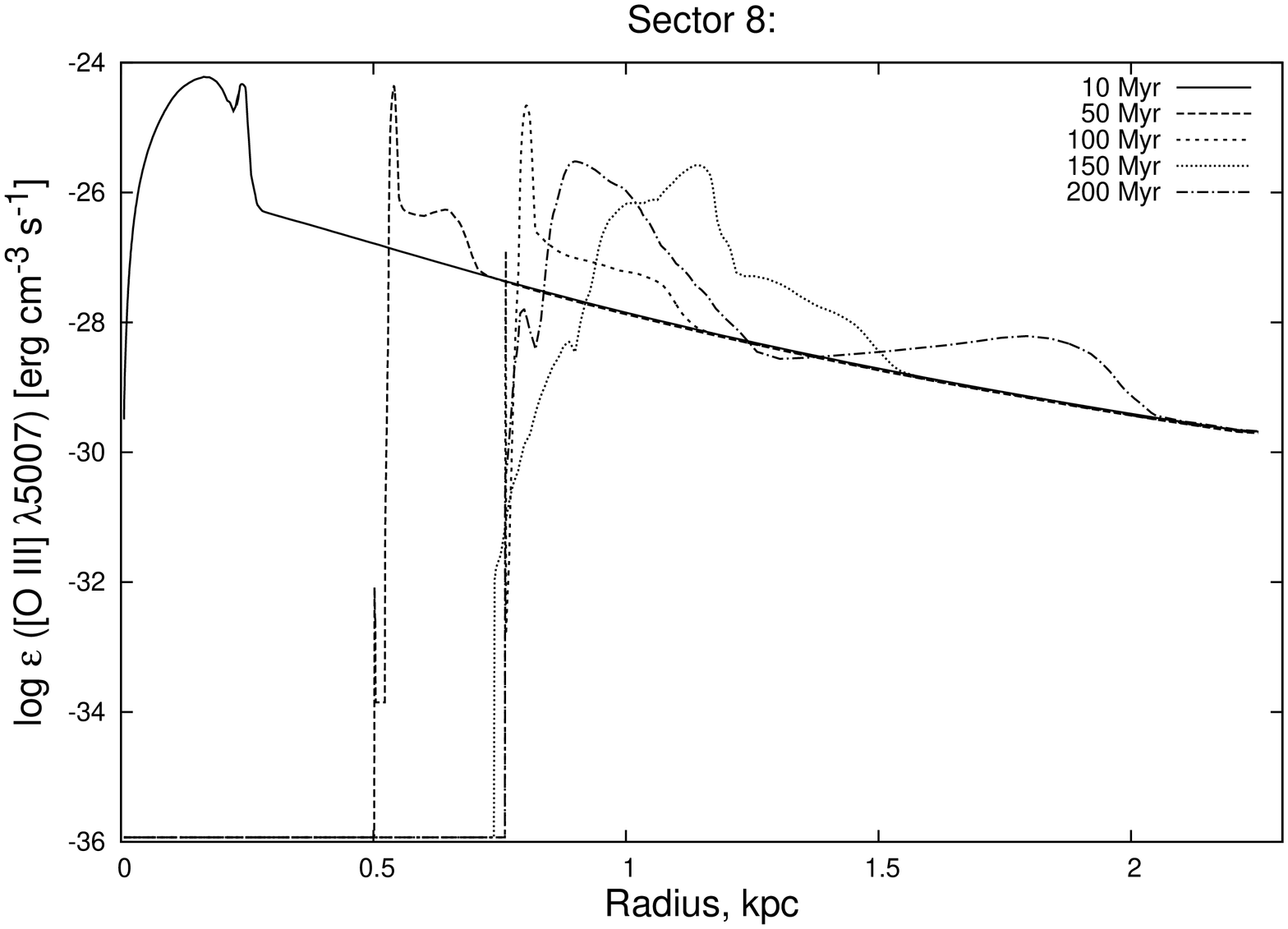}\\
\includegraphics[width=6cm,angle=-90]{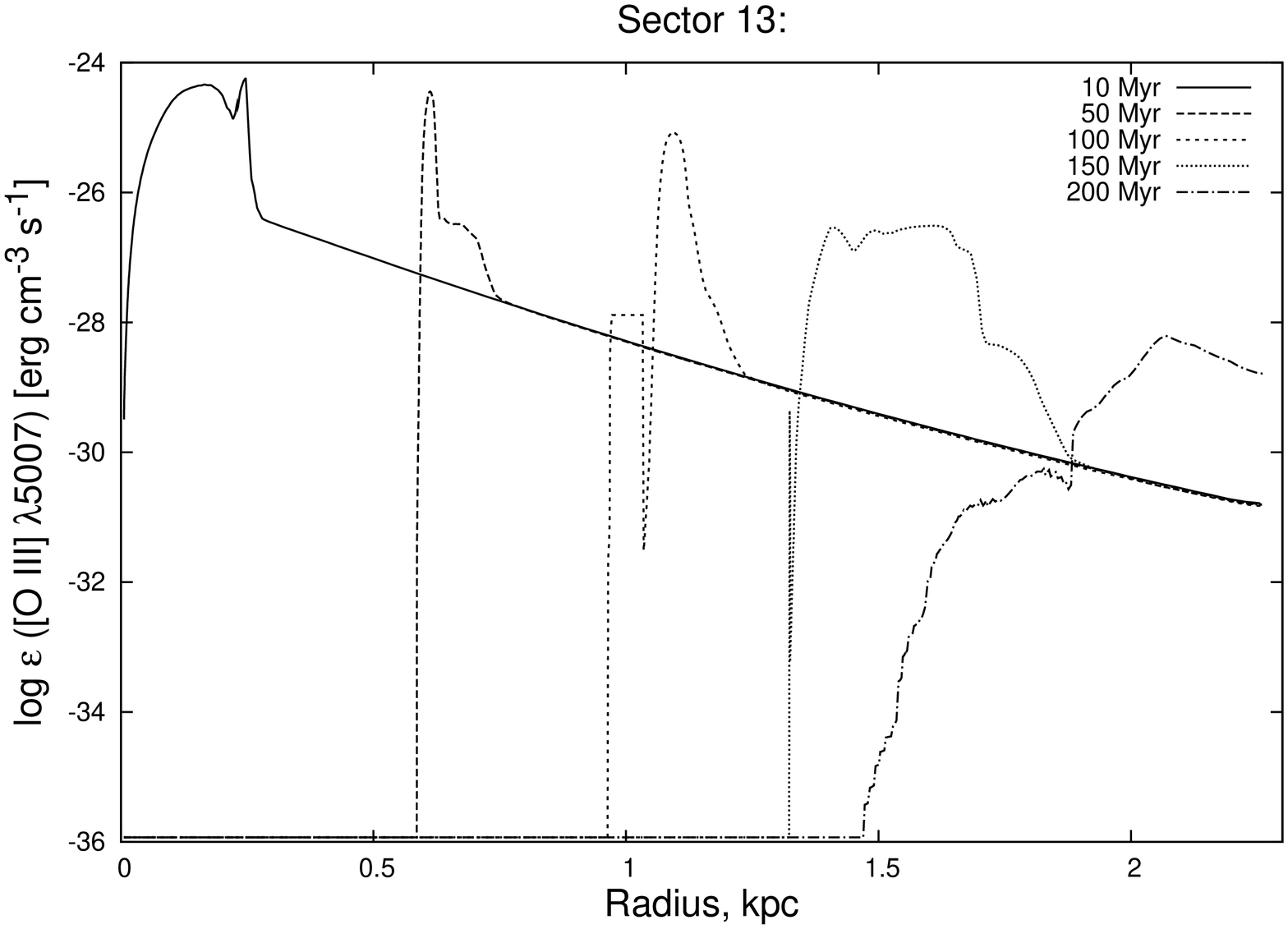}
\includegraphics[width=6cm,angle=-90]{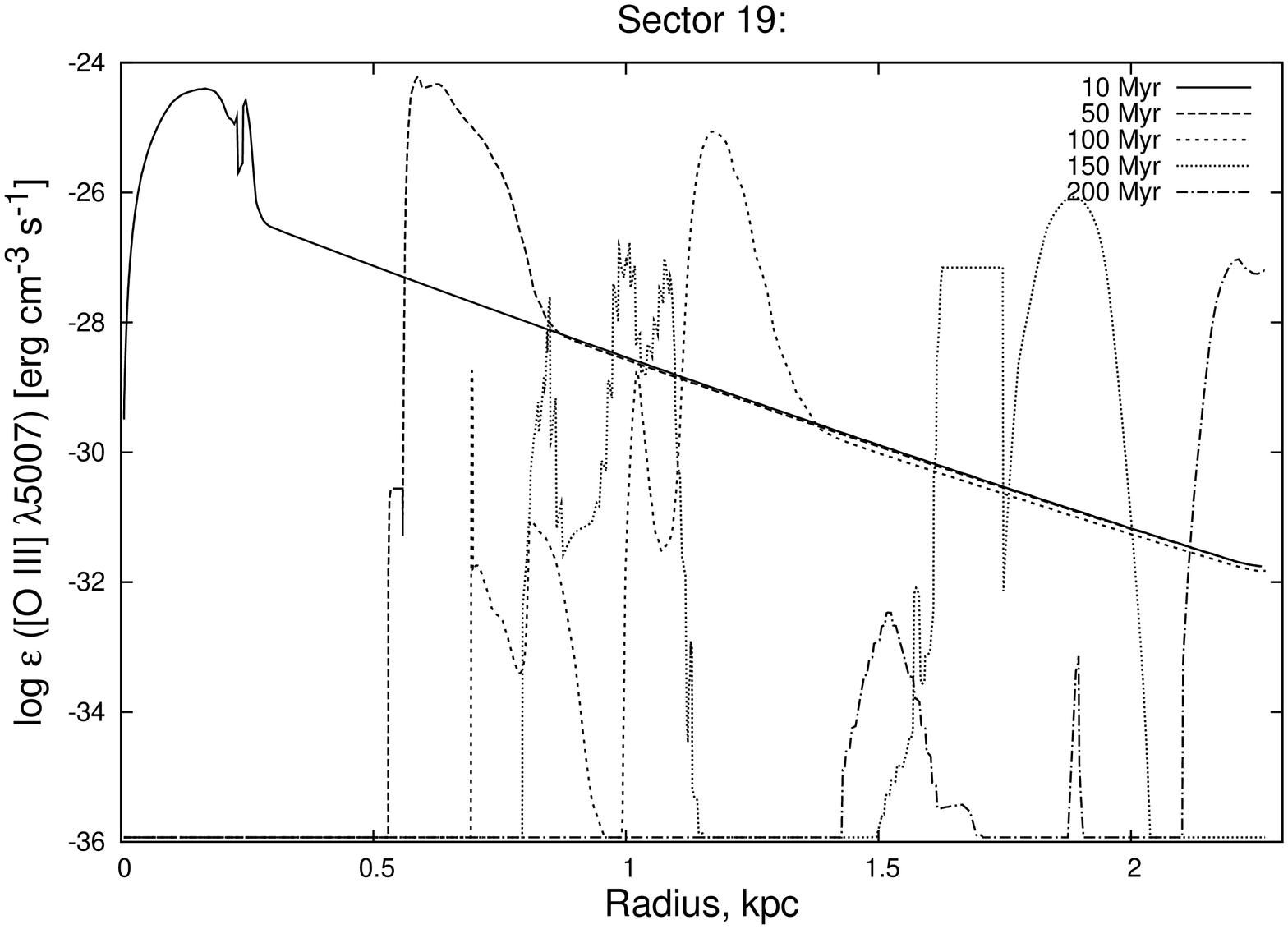}
\caption{Evolution of the radial distribution of $\lambda$5007[O III]
  emissivities for selected sectors and ages.}
\label{fig:5007OIII}
\end{figure*}

In order to investigate where in the nebula the recombination as well
as collisionally excited (forbidden) emission lines arise, we have
plotted the radial distributions of line emissivities (see Figures
\ref{fig:HaHeII} -- \ref{fig:5007OIII}).  We have chosen to plot only
four representative sectors and not all 20 in order to make the plots
clearer.  In particular we have chosen the plane sector (sector 1),
two central sectors (sectors 8 and 13) and a sector close to the axis
(sector 19).  We have not chosen sector 20 because axysimmetric
hydrodynamical simulations (such as our simulations) unavoidably show
some numerical noise along the symmetry axis.  In the left panels of
Figure \ref{fig:HaHeII}, the radial distribution of the \Ha
emissivity (as usual, separated into angular sectors) is shown.  This
line emissivity behaves quite similarly to the electron density (see
Figure \ref{fig:NeTe}) as expected, because this
recombination line has a strong dependence on electron density and a
very weak dependence on the electron temperature. For the first seven
sectors the emissivities drop at the ionization front, whereas for
high-number sectors the external regions of the galaxy are emitting
significantly in \Ha.  This is because the optical thickness of
the gas is lower and the radiation field (mainly due to Lyc photons
from the central SF region) keeps the temperature until $R\sim 1 -
2.5$ kpc relatively high, with $T_e$ ranging between 4000 and 32000 K.
The optical thickness for Lyc photons is not high, because the
intervening gas has on average relatively low densities (of the order
of 0.1 cm$^{-3}$).  This figure is very low in comparison with typical
densities in conventional HII regions, where $n_H\ =\ 1 - 300\
cm^{-3}$.  Notice that in spite of the good spatial resolution of our
hydrodynamical simulations (4 pc), this is still not enough to resolve
HII regions properly (let alone to resolve compact and ultra-compact
HII regions) and this is clearly a limitation of our model, as the
leakage of Lyc photons in our results is perhaps higher than it should 
be (see Sect. \ref{sec:fesc}).  High-resolution multi-phase simulations
are currently being tested, in order to remedy to this situation.  In
this work, for illustrative purposes, we have decided to make a
simpler (and less physically justifiable) hypothesis, namely that a
clumpy phase is present in the galaxy at spatial scales that can not
be solved by our hydrodynamical code.  Therefore, we add this clumpy
phase only during the MPhM post-processing (see next Section for more
detail).

The behaviour of H$\beta$ and HeI$\lambda$4471 as well as other HeI
lines emissivities are also quite similar to the \Ha.  The emissivity
of HeII$\lambda$4686 shows the peaks in 'wall' sectors, corresponding
to maxima in the electron density distribution.  Again, for sector 1,
this emissivity drops to almost zero at the ionization front.  This is
because the ionization potential of He$^+$ is 4Ryd.  The 'wall'
absorbs the photons beyond 4Ryd. Therefore, the fraction of the
ionizing photons beyond the 'wall' is very small.  Thus, at the inner
edge of the 'wall' one has frequently He$^{++}$/He$>$He$^+$/He, but
the He$^{++}$/He fraction drops faster than the He$^+$/He one.  The
emissivity of HeII$\lambda$4686 line is only caused by recombination
of He$^{++}$. In the wind region (SWR) He$^{++}$/He$\gg$He$^+$/He,
because the high temperature here prevents the recombination so that
there does no significant emissivity in HeII$\lambda$4686 line emerge.

In order to illustrate the behaviour of forbidden line emissivities we
show in Fig. \ref{fig:5007OIII} the radial distribution of
[OIII]$\lambda$5007.  This line, like other forbidden lines, is
sensitive to the electron temperature.  The highest temperature values
occur inside the superwind cavity region.  Nevertheless, the
[OIII]$\lambda$5007 emissivity is very weak, because at such high
temperatures the abundance of O$^{++}$ ions is very low (see next
Section).  As expected, for sector 1, characterized by the presence of
an outer ionization front, the [OIII]$\lambda$5007 emissivity drops to
almost zero at the ionization front.

\section{Synthetic spectra, their analysis, and comparison with observations}
\label{sec:abund}
Observed emission line spectra from spatially distributed NG
surrounding the SF region are usually obtained from different aperture
positions \citep[see e.g.][KS97]{KS97}.  Across these apertures,
abundances of various chemical elements can be derived by different
diagnostic methods.  The aim of this section is to compare the
chemical abundances directly obtained from ChDSs with the ones
obtained after applying diagnostic methods to the MPhM output files.
In this way we try to assess the validity of various diagnostic
methods, as well as the correctness of our methods and procedures.
In fact, as illustrated in the previous Sections, our
combined ChDS+MPhM simulations allowed us to produce 3D maps of
emission line and continuum emissivities, which we illustrated by
means of radial variations across various angular sectors.  On the
other hand, by means of ChDSs alone we have the spatial distribution
of the abundance of nine chemical elements (H, He, C, N, O, Mg, Si,
Fe), as a function of time.  We thus have all required information to
test the diagnostic methods.

\begin{figure*}
\centering
\includegraphics[width=0.5\linewidth]{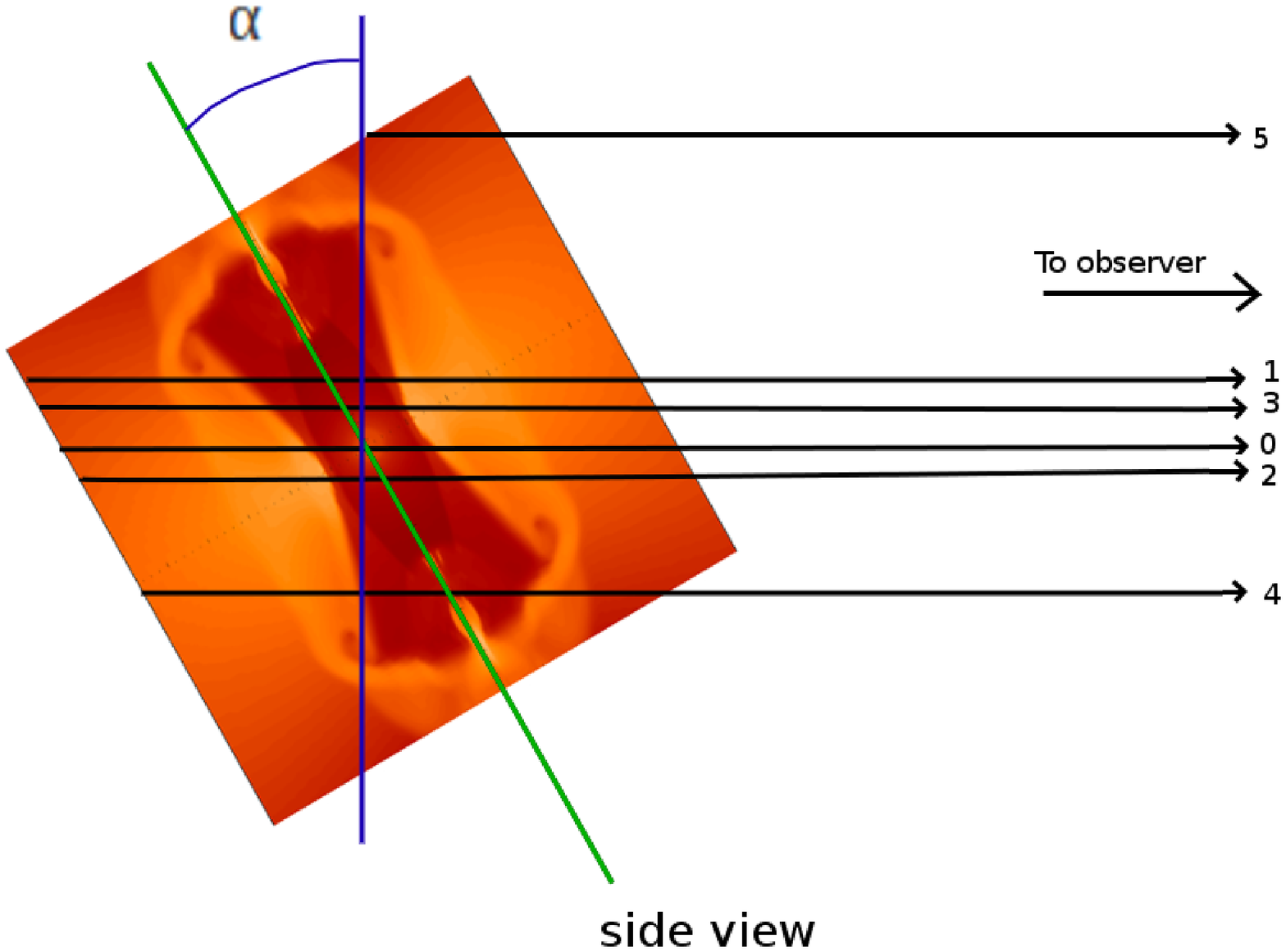}
\includegraphics[width=0.43\linewidth]{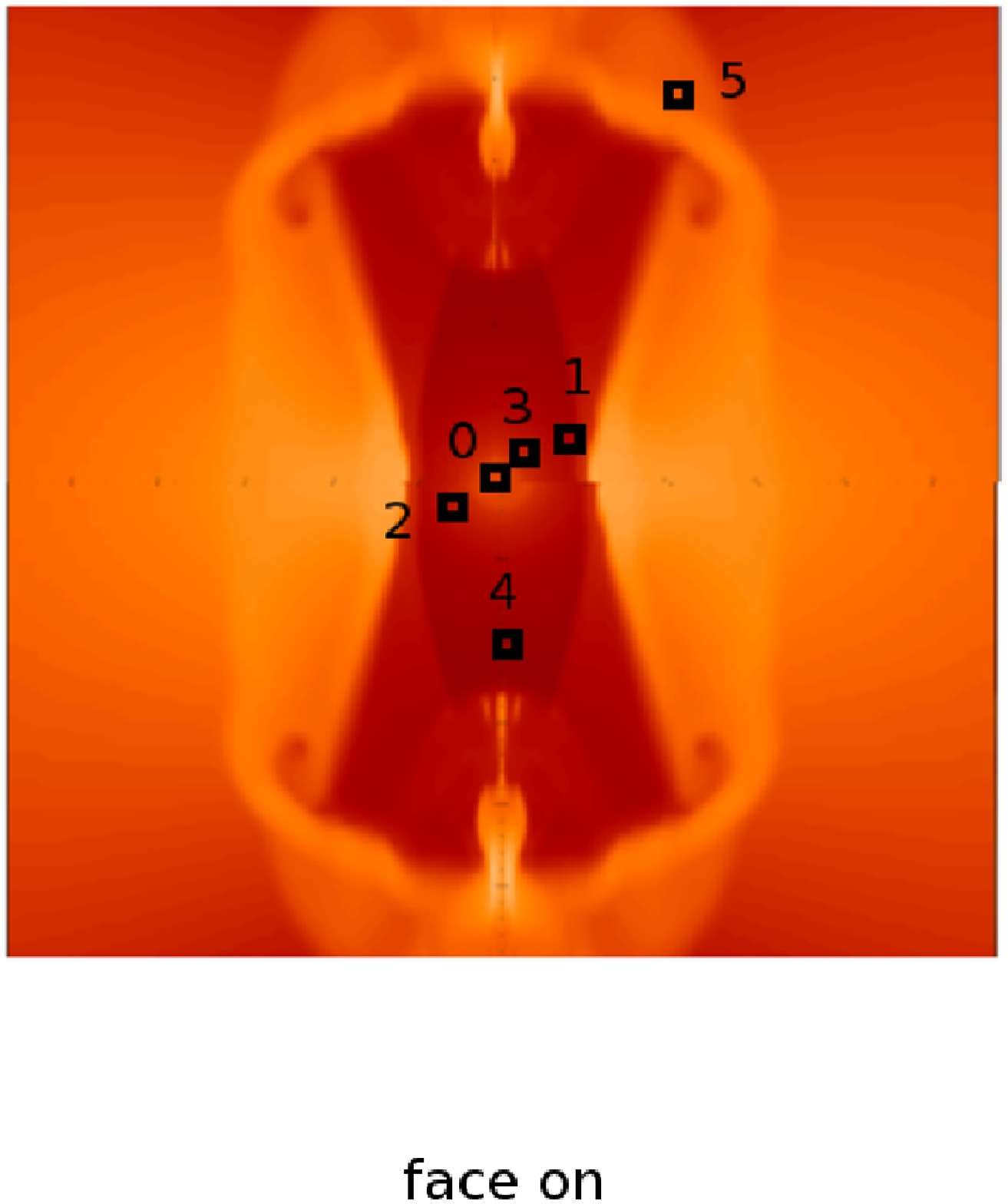}
\caption{The definition of aperture cuts. The central aperture is
  labeled by zero.  $\alpha$ is the angle between the symmetry axis
  and the plane normal to the observer sight line. The left panel is
  the side view of the object.  The right panel is the front view.
  Here, the size of the aperture can be appreciated.}
\label{fig:apertures}
\end{figure*}

In this Section we shortly describe the procedure to calculate the
synthetic emission line spectra from different aperture positions
using emissivity maps.  We firstly define non-central (i.e. along
lines-of-sight which do not pass through the center of the DG)) and
central aperture positions (see Fig.  \ref{fig:apertures}).  Synthetic
spectra are calculated by integrating the intensity maps across each
aperture (see next subsection for more details).  Synthetic spectra
are in turn used to determine the oxygen abundances using two
well-known and widely used diagnostic methods, namely the two-zones
$T_e$-method \citep[see][]{Te_method,Te_method2} and the
$R_{23}$-method.  The latter is based on calibrating relationships
O/H-$R_{23}$ from \citet{McCaugh}.  Then we present the comparison of
the obtained oxygen abundances with the ones from the ChDS for the
corresponding aperture.

\subsection{Synthetic spectrum calculation}

To calculate the emission line spectrum using emissivity maps we
developed the 3D code {\it DiffRaY}. This code integrates the fluxes
in emission lines over the solid angle that is defined by the aperture
position relative to the observer position.

We consider each aperture not as an infinitesimally thin needle, but
as a geometrical figure having a section of 49~pc $\times$ 49~pc.  
This section area is further subdivided in $n\times n$ smaller
volumes.  The subdivision process is iterative and $n$ represents the
stage of the iteration.  $n$ is equal to four in the first iteration.
At each stage of iteration, the radiative fluxes are integrated across
each subdivision of the aperture.  Among different possible numerical
integration techniques \citep[see][]{Numrec}, we have chosen the
trapezoid method, as it is the optimal choice for this kind of
problems \citep[see][]{BuhajenkoMelekh}.  The code repeats this
procedure, increasing $n$ by 2 at each step, and iterates until
convergence, i.e. until the differences between line intensities at
two different stages of iteration are less than 2 \%.

We tested this code by comparing the radial line fluxes obtained from
this 3D integration program with the ones obtained by CLOUDY.  In
Table \ref{Tab:Test} the ratios between our results and CLOUDY's ones
are given.  It can be seen that these ratios are always very close to
unity, with deviations of $\sim$ 1.7\% at most.

\subsection{Synthetic spectra from different aperture positions}
\label{subs:APs}

In order to test the ability of the two-zones $T_e$-method
\citep{Te_method,Te_method2} and $R_{23}$-method \citep{Pagel79} to
determine the oxygen abundances in galaxies, we calculated synthetic
emission line spectra using MPhM emissivity maps, as described above.

The angle $\alpha = 30^o$ (see Fig.\ref{fig:apertures}) characterizing
the inclination of the galaxy relative to observer was adopted.

\begin{table}
\begin{center}
  \caption{The ratios between radial fluxes, calculated by our 3D
    integration program DiffRaY and the ones obtained by CLOUDY for
    the sectors 1, 8, 13, and 19.}
\label{Tab:Test}
    
\begin{tabular}{lcccc}
\hline
Line & Sector 1    &Sector 8& Sector 13& Sector 19\\
\hline              
H$\beta$                  &0.985& 1.002 & 1.002 & 0.985\\
\Ha 		  &0.985& 0.994 & 0.997 & 0.995\\
Ly$\alpha$ $\lambda$1216  &0.984& 1.011 & 0.991 & 0.983\\
$\rm [OII]\lambda$3729  &0.998& 1.003 & 0.999 & 0.990\\
$\rm [OII]\lambda$3726 &0.999& 0.992 & 0.998 & 0.998\\
$\rm [OIII]\lambda$5007 &0.984& 0.990 & 0.996 & 1.003\\
$\rm [OIII]\lambda$4363 &1.000& 0.994 & 1.000 & 0.979\\
HeI $\lambda$4471 &0.998& 0.991 & 0.989 & 0.993\\
HeI $\lambda$5876 &0.994& 0.991 & 0.995 & 1.004\\
HeI $\lambda$6678 &0.995& 0.996 & 1.006 & 1.002\\
HeI $\lambda$7065 &0.987& 0.991 & 0.993 & 0.994\\
$\rm [SII] \lambda$6716 &0.988& 1.002 & 1.008 & 0.984\\
$\rm [SII] \lambda$6731 &1.004& 0.992 & 0.993 & 1.000\\
$\rm [NII] \lambda$6548 &0.989& 0.995 & 1.004 & 1.001\\
$\rm [NII] \lambda$6584 &0.989& 0.996 & 1.004 & 1.000\\
\hline
\end{tabular}
\end{center}
\end{table}

\subsection{Comparison of modelled and observed emission line spectra}
\label{subs:intratios}

\begin{figure}
\centering
\includegraphics[width=6cm,angle=-90]{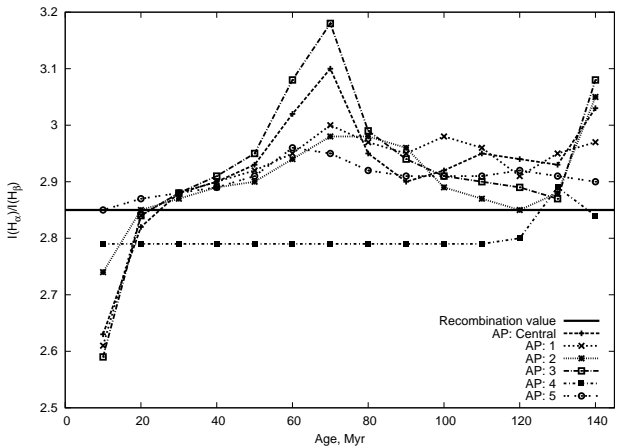}
\caption{The evolution of H$_\alpha$/H$_\beta$ for six synthetic
  aperture positions (AP).}
\label{fig:HaHb}
\end{figure}

The aim of this subsection is to compare the calculated relative line
intensities with the observed ones. For this
  comparison with observations we use the sample of selected observed
  spectra from \citet{Kehrig2004}.  From this sample, we selected
  galaxies with H$\alpha$/H$\beta$ between 2.6 and 3.2, because our
  models results are in this range (see Figure~\ref{fig:HaHb}).  The
  resulting observational ranges for the intensities of important
  emission lines are given in column 2 of Table \ref{Tab:ObsCompar}.
  Of course, we do not expect a perfect match between theoretical
expectations and observations, because this paper is not intended for
this purpose, but model emission line intensities should be the of the
same order as observed ones.

In Table 2 and in what follows, we label with M1 the standard model,
i.e. the analysis done starting from the results of the ChDS.  Model
M2 refers to a modification of the ChDS results, in which a very thin
shell of enhanced density is artificially added (see below for more
detail).

\begin{table}
\begin{center}
  \caption{Comparison of relative line intensities observed at
      galaxies, selected from the sample of \citet{Kehrig2004} with
    derived ones of models M1 and M2 (see text).}
\label{Tab:ObsCompar}
\begin{tabular}{lcrr}
\hline
Relative intensity 			& Observed 	& M1    & M2  \\
                                        &  range   	&       &     \\
\hline
${\rm [OIII]}{\lambda 5007}/{\rm H\beta}$ &  0.58 .. 7.84  &2.08 &3.47\\
${\rm [OIII]}{\lambda 4363}/{\rm H\beta}$ &  0.03 .. 0.15  &0.03 &0.03\\
${\rm [OII]}{\lambda 3727}/{\rm H\beta}^*$&  0.40 .. 7.19  &0.18 &1.95\\
${\rm [OIII]}{\lambda 5007}/{\rm [OII]}{\lambda 3727}$ & 0.17 .. 4.8 & 11.6 &1.78\\
${\rm [SII]}{\lambda 6716}/{\rm H\beta}$  & 0.05 .. 0.66  &0.02 &0.23\\
${\rm [SII]}{\lambda 6731}/{\rm H\beta}$  & 0.04 .. 0.76  &0.02 &0.16\\
${\rm H\alpha}/{\rm H\beta}$  	          & 2.63 .. 3.18  &2.95 &2.93\\
${\rm [NII]}{\lambda 6584}/{\rm H\alpha}$ & 0.02 .. 0.29  &0.001 &0.025\\
${\rm [SII]}{\lambda 6716}/{\rm H\alpha}$  & 0.02 .. 0.21 &0.007 &0.08\\
${\rm [SII]}{\lambda 6731}/{\rm H\alpha}$  & 0.02 .. 0.25 &0.007 &0.06\\
$\rm [SII]({\lambda 6716}+{\lambda 6731})/{\rm H\alpha}$ & 0.03 .. 0.46 &0.01 &0.13\\
\hline
\end{tabular}
$^*{\rm [O II]}{\lambda 3727}/{\rm H\beta}={\rm [O II]}(\lambda 3726+\lambda 3729)/{\rm H\beta}$\\
\end{center}
\end{table}

\begin{figure*}
\centering
\includegraphics[width=6cm,angle=-90]{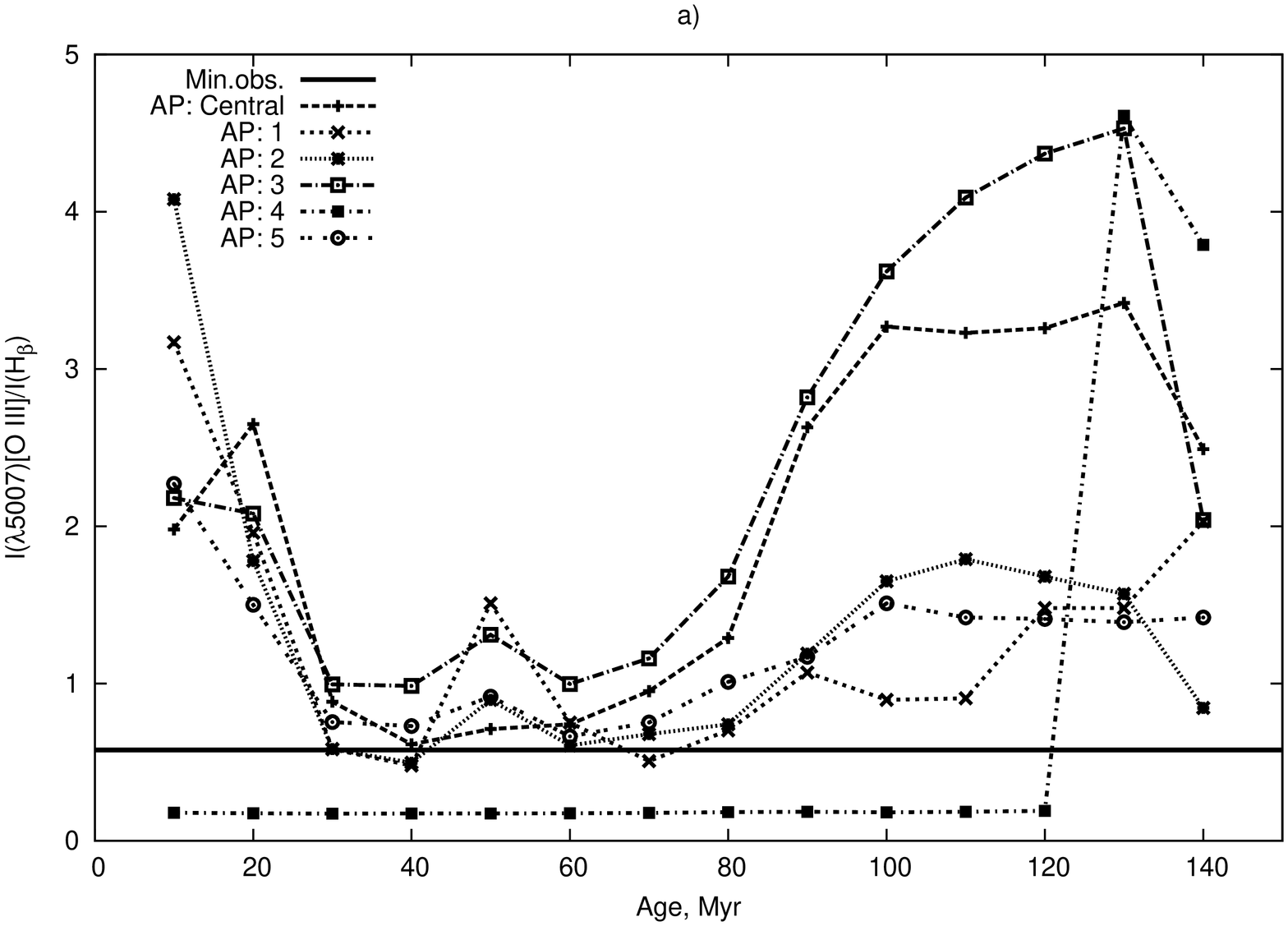}
\includegraphics[width=6cm,angle=-90]{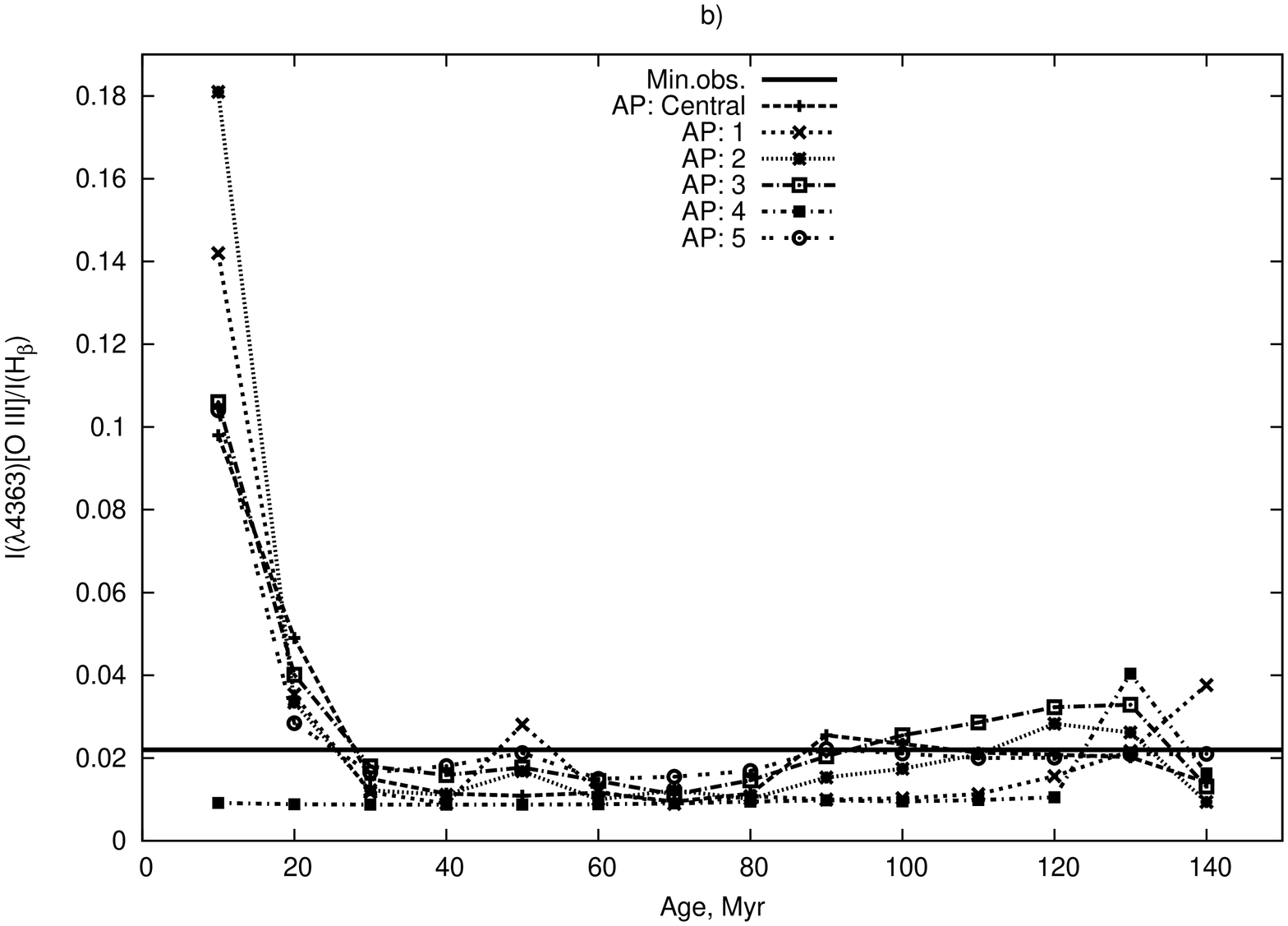}
\includegraphics[width=6cm,angle=-90]{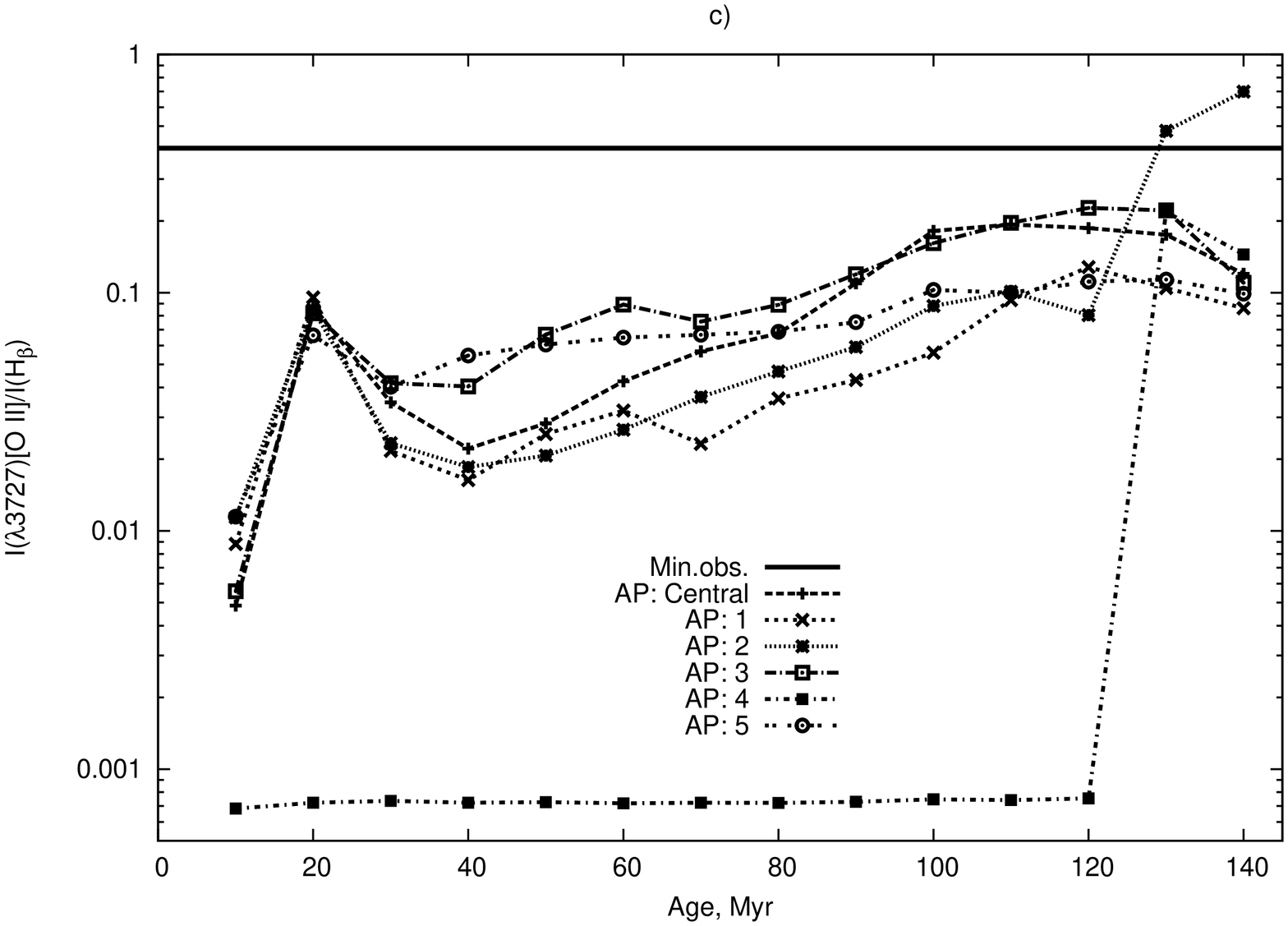}
\includegraphics[width=6cm,angle=-90]{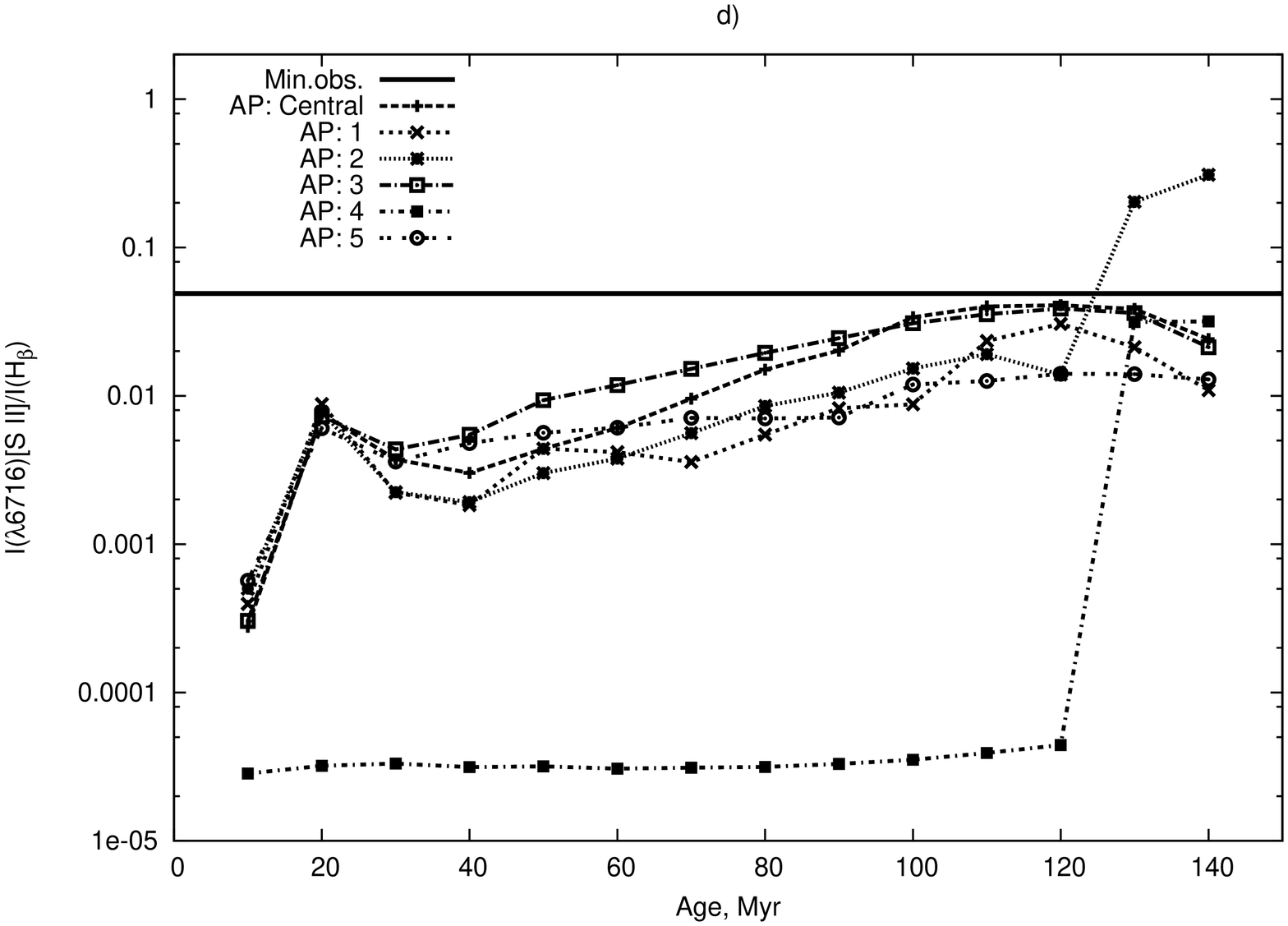}
\\
\caption{Some important line intensities, obtained for six synthetic
  aperture positions (AP), as a function of time: a)~[OIII]$\lambda
  5007$ -- one of the main coolers of nebular gas, also used in
  diagnostics for oxygen abundance determination; b)[OIII]$\lambda
  4363$-- important line for electron temperature determination;
  c)~[OII]$\lambda 3727$ -- it is a sum of [OII]$\lambda 3726$ and
  [OII]$\lambda 3729$ line intensities, that are important to
  determine the oxygen abundance; d)~[SII]$\lambda 6716$ -- the ratio
  of this line to [SII]$\lambda 6731$ is used in diagnostic for
  electron density determination.  The intensities are normalized to
  $I_{\rm H\beta}$.  To compare with observations the minimal value,
  obtained from observed emission line spectra of galaxies,
    selected from sample by \citet{Kehrig2004} are shown by solid
  lines (Min.obs.) in each panel.}
\label{fig:LineIntensAper}
\end{figure*}

In Figures \ref{fig:LineIntensAper} the time dependences of the
relative line intensities (relative to H$\beta$) of important emission
lines, obtained for various synthetic aperture positions (see
Fig.~\ref{fig:apertures}) are shown.  It can be seen that our models
predict [OIII]$\lambda 5007$ (see Fig.~\ref{fig:LineIntensAper}a))
line intensities higher than the observed minimum or very close to the
one for most of APs.  Only in the case of AP~4 intnesity of this
  line is higher than the observed minimum after 120~Myr.
The relative intensities of [OIII]$\lambda 4363$, [OII]$\lambda 3727$ and
[SII]$\lambda 6716$  (see Fig.~\ref{fig:LineIntensAper}b--d) ) for most of ages and APs 
are not able to reproduce the observed minimum.

To further simplify the comparison of the model results with observed
ones, we select the relative intensities obtained as a function of
radial distance of sector 1 at 140 Myr.  The resulting intensities are
given in column M1 of Table \ref{Tab:ObsCompar}. This angular sector
has the same problems shown by the aperture positions above, i.e. it
severely underpredicts the [OII]$\lambda 3727$ and [SII]$\lambda 6716$
emission line intensities.
We also compare the intensities of ${\rm H\alpha}/{\rm H\beta}$, 
of [NII]$\lambda 6584/{\rm H\alpha}$, and of the sum of
[SII]$\lambda 6716/{\rm H\alpha}$ and [SII]$\lambda 6731/{\rm
  H\alpha}$ with observed ones.  It can be seen that only the
  values of ${\rm [OIII]}{\lambda5007}/{\rm H\beta}$, ${\rm
  H\alpha}/{\rm H\beta}$ and ${\rm [OIII]}{\lambda4363}/{\rm H\beta}$
are within the observed ranges.  The other line ratios are
underestimated in model M1.

\begin{figure*}
\centering
\includegraphics[width=6cm,angle=-90]{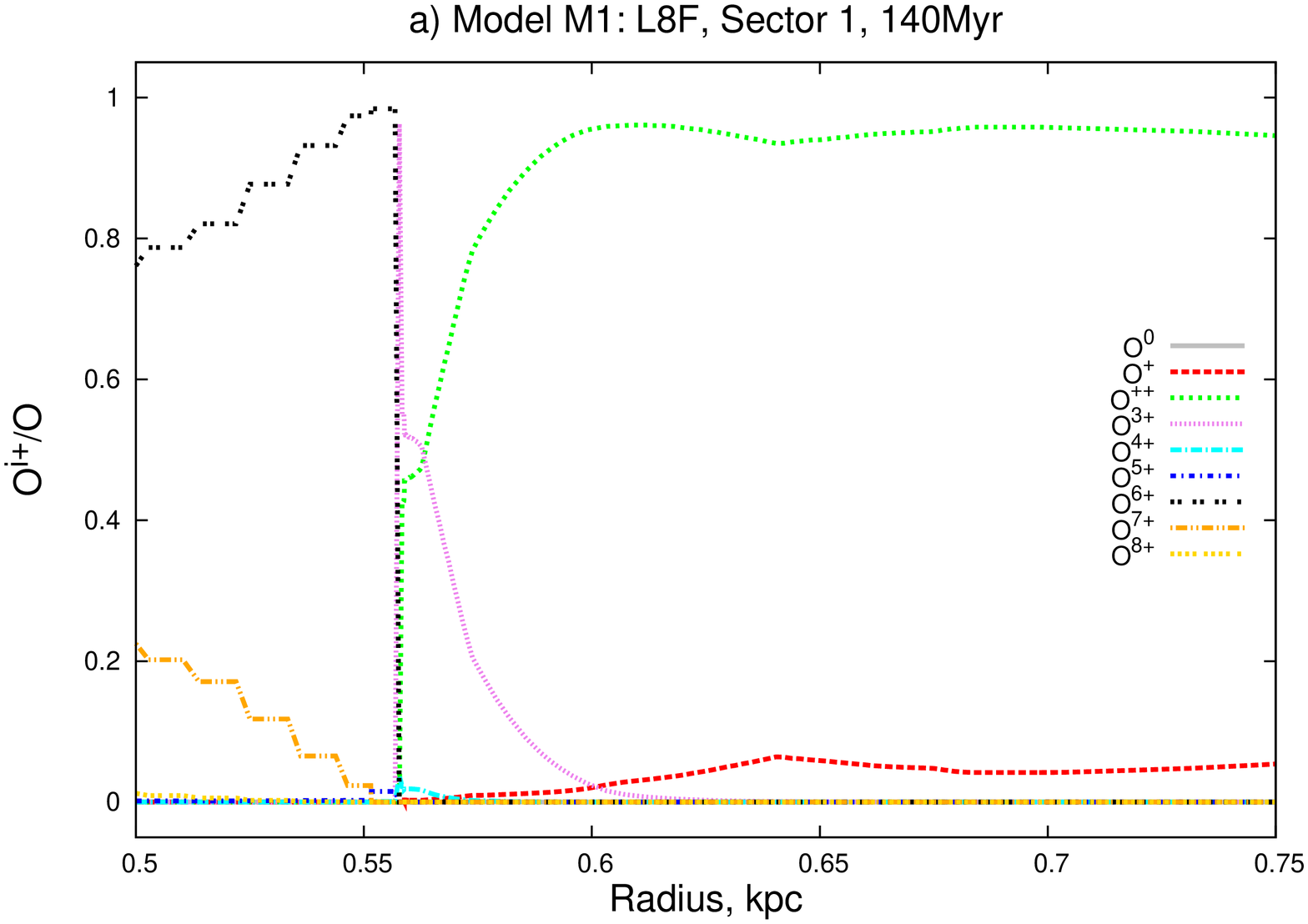}
\includegraphics[width=6cm,angle=-90]{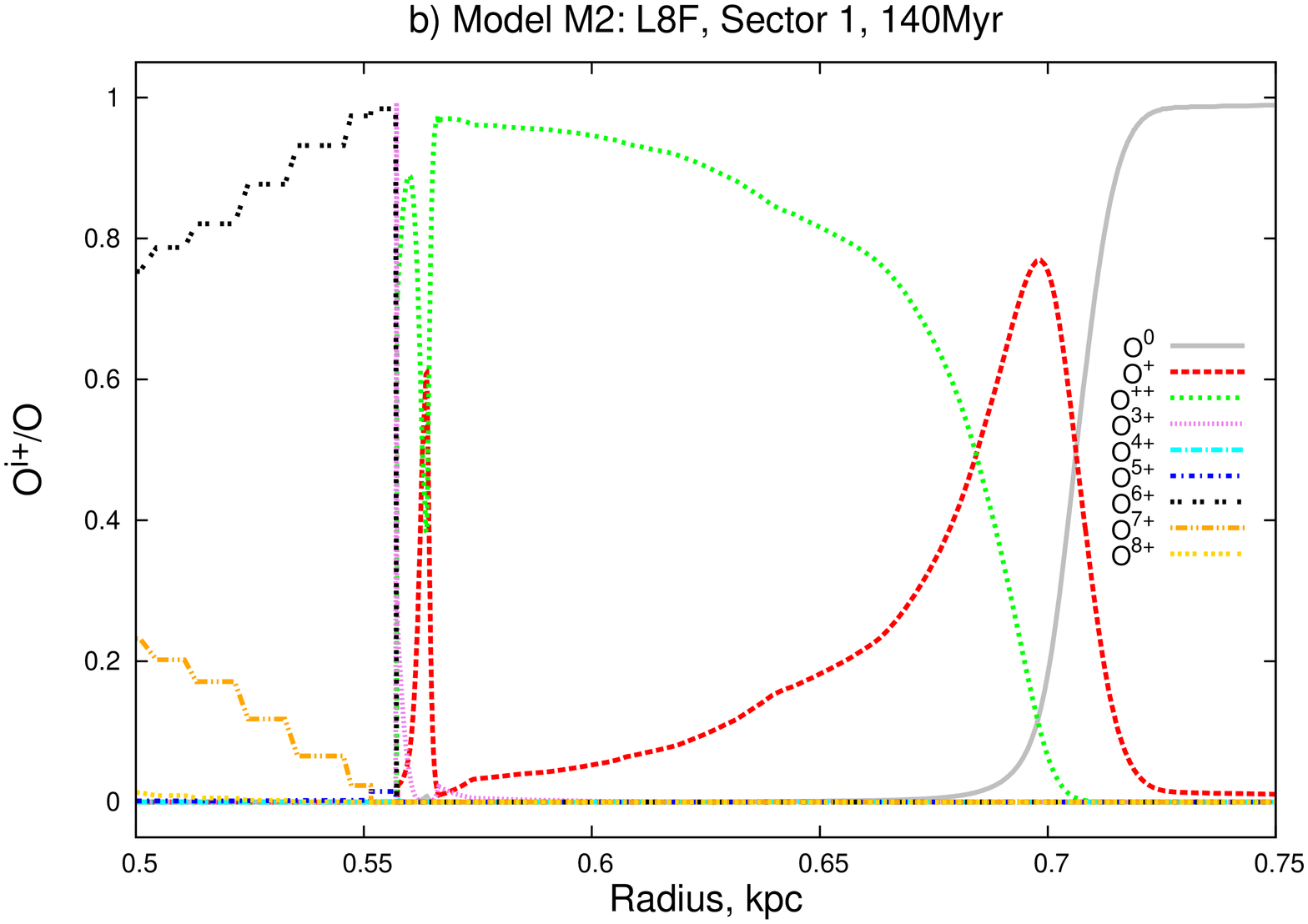}
\\
\includegraphics[width=6cm,angle=-90]{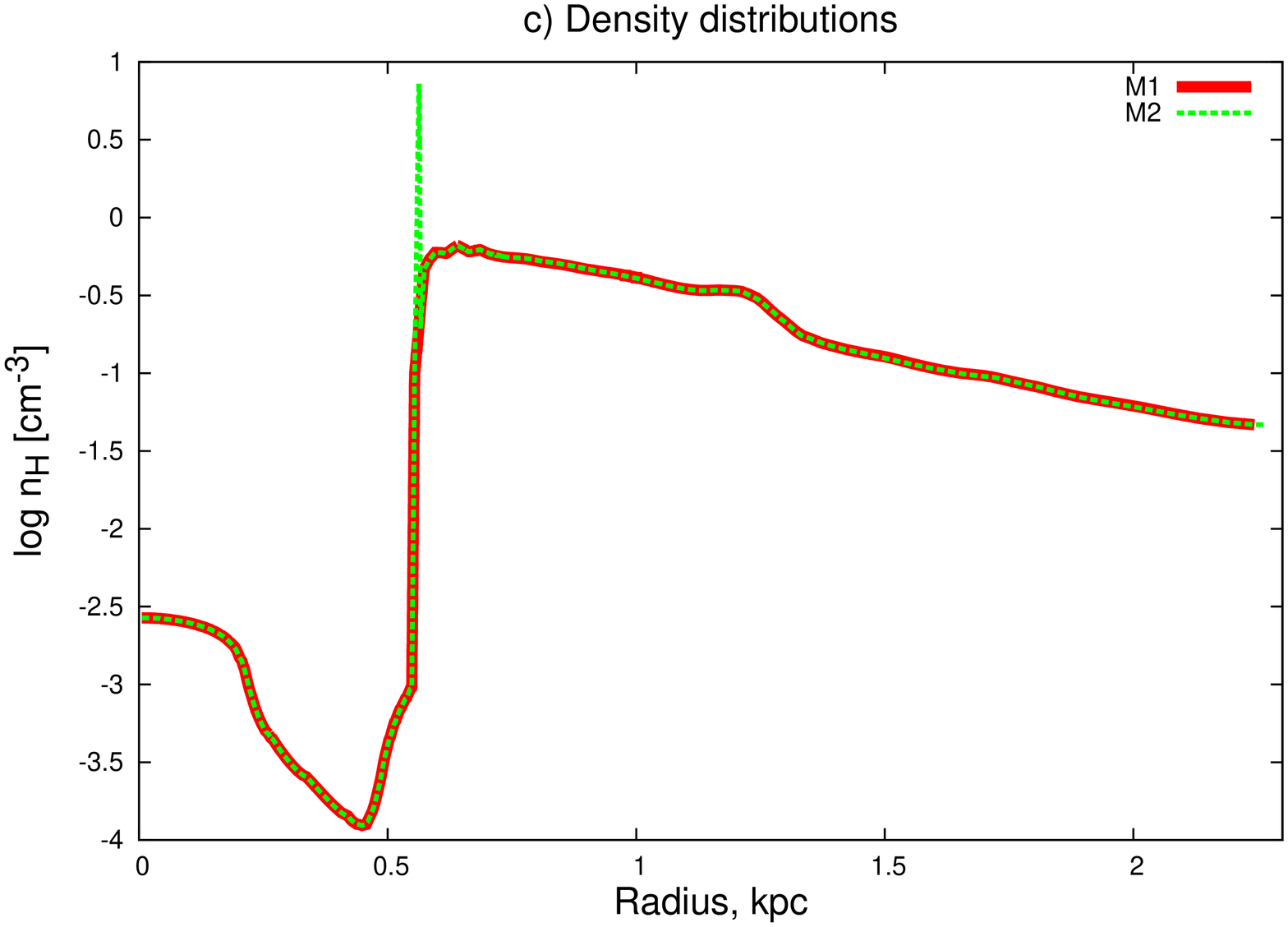}
\includegraphics[width=6cm,angle=-90]{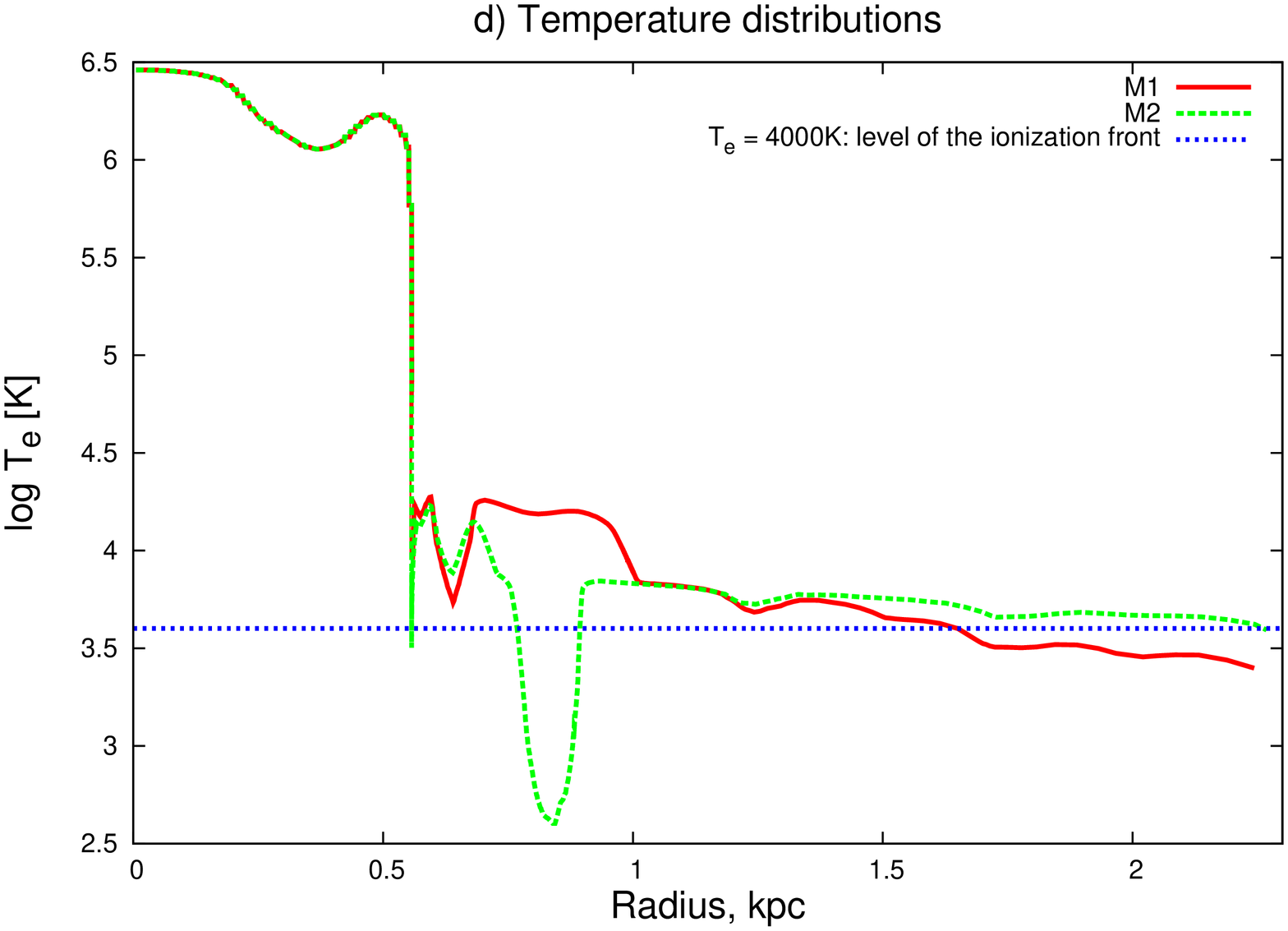}
\caption{Radial distribution of the oxygen ionization fractions,
  hydrogen densities and electron temperatures in Sector 1 at 140 Myr
  in models M1 and M2.  Model M2 is characterized by the addition of a
  thin shell of enhanced density in 'wall' with a peak density of $7
  cm^{-3}$ (for comparison, the standard density obtained from ChDS at
  this position is about $0.2cm^{-3}$).  }
\label{fig:OxygenM}
\end{figure*}

The difficulty in reproducing the correct [OII] and [SII]
intensities boils down in the end to the relatively low gas density
across most of the computational volume; a limitation of the model we
have already outlined in the previous Section.  Due to this low
density, the temperature beyond the 'wall' in most cases is relatively
high.  This prevents the formation of a significant fraction of ${\rm
  O^+}$ ions.  In this part of the galaxy, the oxygen is in fact
mostly doubly ionized and this explains the reasonable agreement
between [OIII] intensities and observations.  To illustrate this
point, the various ionization stages of oxygen across the wall at
t=140 Myr are shown in Fig. \ref{fig:OxygenM}, upper-left panel.  We
have chosen the angular sector 1 (i.e. the one adjacent to the plane
of the galaxy, see previous Section), between 0.5 and 0.75 kpc because
the radial distribution of the various ionization stages can be better
appreciated.

\begin{figure}
\centering
\includegraphics[width=6.2cm,angle=-90]{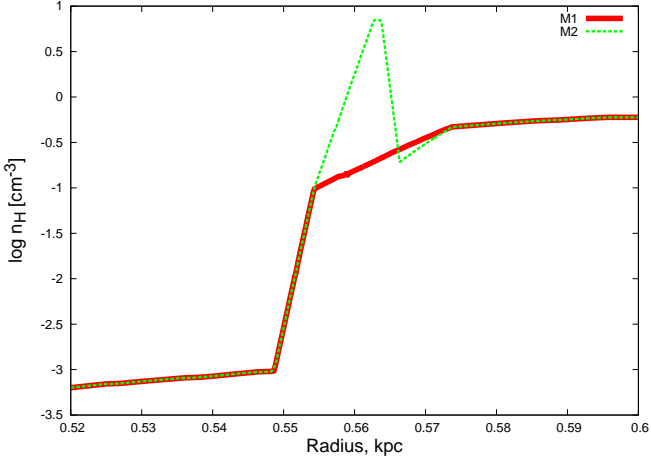}
\caption{Modification of density distribution in model M2 in
comparison with standard model M1 calculated at 140Myr.}
\label{fig:DensM1M3}
\end{figure}

In order to address the problem of the lack of very dense regions in
our computational volume, we decided to introduce one dense shell of
enhanced density (TDS -- thin dense shell).  This TDS represents
  a thin density enhancement, due to a shock, impossible to resolve
  with the currently adopted numerical resolution. The shell extends
for 10 pc (see Fig. \ref{fig:DensM1M3}) and reaches a peak density of
$n_H = 7cm^{-3}$. This peak density (not the entire shell; just the peak) 
extends for 1 pc.  The M1 density at the peak position is instead about $0.2$ cm$^{-3}$.  
  With such a background density, the TDS peak density would thus
  represent a density enhancement of $\sim$ 35 in a shock.  For a
  strong isothermal shock, this would correspond to a Mach number of
  $\simeq$ 6 \citep[for instance see the textbook of][]{Draine2011}, perfectly 
  reasonable under the studied conditions.

To illustrate the effect of the TDS presence we have chosen again
Sector 1, because it is the sector adjacent to the plane of the
galaxy, where the presence of a TDS is more likely.  The TDS inherits
the chemical composition of the underlying material.  We calculated
models of Sector 1 that differ by the position of this TDS: at the
wall; midway between the wall and the outer radius of the considered
computational box ($R_{out} \approx$ 2.3kpc); and very close to
$R_{out}$.  Only in the case of a TDS positioned in the wall we
obtained relative intensities in agreement with observations. The
results of this model, labelled M2, are shown in Table 2.  Also, in
Figure \ref{fig:OxygenM} (upper-right panel) the various ionization
stages of oxygen for M2 are shown.  The TDS produces a significant
increase of the ${\rm O^+/O}$ ratio at a distance larger than 0.55
kpc.  Now, the radial profile of of ${\rm O^+}$ in model M2 is more
complicated (there are two maxima).  This is enough to significantly
increase the [O II] and [S II] emission.  The disagreement with the
observations disappears for M2 (see again Table 2).  The values of
$I_{\rm H\alpha}/I_{\rm H\beta}$, $I_{\lambda 6584}{\rm [NII]}/I_{\rm
  H\alpha}$ as well as $(I_{\lambda 6716}{\rm [SII]}+I_{\lambda
  6731}{\rm [SII]})/I_{\rm H\alpha}$ are within the observed range
too.

The combination of $I_{\lambda 5007}{\rm [O III]}/I_{\rm H\beta}$ and
$I_{\lambda 6584}{\rm [NII]}/I_{\rm H\alpha}$ allows us to place our
models in the so-called BPT diagram \citep{BPT1981}, the famous
seagull-shaped diagram able to distinguish between HII regions due to
star formation and AGNs.  For model M2 $(\log\ I_{\lambda 6584}{\rm
  [NII]}/I_{\rm H\alpha}\ ,\ \log\ I_{\lambda 5007}{\rm [OIII]}/I_{\rm
  H\beta})=(-1.14,0.54)$.  Comparing these values with fig. 5 of
\citet{BPT1981}, it is clear that model M2 is well within the HII
regions.  Our results obey pretty precisely eq. 8 of the Baldwin et
al. paper.

In the SWR, the oxygen is present in ${\rm O^{3+}}$, ${\rm O^{4+}}$
and even higher ionization stages, because of the high temperatures.
The density and temperature distributions for models M1 and M2 as well
as the temperature level of the ionization front are shown in Figure
\ref{fig:OxygenM} (panels c and d).  Because of the TDS, the
temperature drops significantly below the ionization front, at
distances of the order of 0.7--0.9 kpc. At larger distances, the
temperature increases again, because of the reduced density.  By means
of the TDS, we are thus able to create a layer of neutral gas between
two ionized regions.

In the M1 model, the oxygen at large radii exists mainly in the
${\rm O^{2+}}$ ionization stage, therefore the ${\rm O^{+}}$ abundance
is too low to reproduce the observed [OII] line intensities.  The
presence of the TDS (model M2) leads to the appearance of two regions
of enhanced by ${\rm O^{+}}$ fractional abundance: the first one is a
thin region at the inner edge of the wall; the second one extends up
to $\approx$ 0.73 kpc.  These ${\rm O^{+}}$ shells are enough to
reproduce the observed [OII] emission in observed range (see the
results of model M2 in Table 2).

Although the results of model M2 are encouraging, it is important to
point out that our results are quite sensitive to the TDS position and
also to the density profile within the TDS.  It must be also
emphasized that, in our models, diffuse ionizing radiation (emitted by
ionized diffuse gas) is calculated in outward only approximation.
Therefore, as already outlined, ionizing photons from neighbor sectors
cannot penetrate.  Therefore, the propagation of the ionizing
radiation extends up to the outer ionization front, that occurs at
$\approx$ 1 kpc.  Such a restriction for the propagation of ionizing
photons does not exists in real galaxies.  We are working on an
extension of the MPhM code to deal with this kind of non-radial
propagation of ionizing photons.   Such extension is
especially necessary to include compact, non-axisymmetric clumps in
the model, and to consider spatially extended and non-spherical
sources of ionizing radiation.

\subsection{Diagnostics of synthetic emission line spectra}
\label{subs:abundances}

Our study aims also at comparing the derived oxygen abundances with
the ChDS ones, by means of two popular diagnostics, the $T_e$-method
and the $R_{23}$-method.  

According to $T_e$-method \citep[see][]{Te_method,Te_method2}, 
for the determination of the oxygen abundances, we also need 
the relative intensities of
[OIII]($\lambda4363$,$\lambda4959$,$\lambda5007$),
and [OII]($\lambda3726$,$\lambda3729$), obtained from the synthetic
emission line spectra and calculated above for different aperture
positions.

However, frequently the emission line $\lambda4363$[OIII] is absent in
real observed spectra, because of low intensity.  In such cases the
so-called calibration methods are used.  In such methods the
relationships between (12~+~log~O/H) and the parameter $R_{23} =
(I_{\rm[OII]\lambda3726+\lambda3729} + I_{\rm[OIII]\lambda4959+
  \lambda5007})/I_{\rm H\beta}$ are determined on the basis of
appropriate calibrations.  We select for our tests the calibration of
\citet[]{McCaugh}.

In Table 3 we compare the oxygen abundances,  the luminosity-weighted (L-W)
and mass-weighted (M-W) with the ones calculated for the six selected synthetic
apertures, using the chosen diagnostic methods, for three different
ages. The behaviour of O/H abundances along sight lines becomes more
complicate with age.  The $T_e$-method reproduce the L-W values within
an error ($\Delta$) of only 0.5~dex for three APs at 10 Myr, for one
AP at 70 Myr and for two APs at 140 Myr.
Larger deviations from L-W results can be explained by the
inhomogeneous distribution of $\rm O^{+}$ and $\rm O^{++}$ ions.
Also, the $R_{23}$-method does not reproduce the oxygen abundance in
AP 5.
In most cases, the $R_{23}$-method overestimates the oxygen abundances
with deviations of more than 0.5~dex, but for some apertures, the L-W
abundances are very close to the ones calculated with the
$R_{23}$-method.

These results show that the considered diagnostic methods cannot be
used to precisely determine the oxygen abundance in DGs due to both
the inhomogeneity of the element abundance and the very low density
($\le$3cm$^{-3}$ in some cases).  The results show also that
the model limitations, already pointed out above, might render our
model galaxies quite different from real galaxies.  In particular,
as already pointed out, due to the low density, the optical thickness
for the emission lines along most of sight lines are low 
(probably lower than in real galaxies).

Also in Table 3 (last two rows) we compare L-W and M-W oxygen
abundances with the ones determined using both $T_e$- and
$R_{23}$-methods for models M1 and M2 at 140 Myr.  It can be seen that
for M1 the difference between the $T_e$ and the $R_{23}$ method
amounts to 0.4~dex.  The differences between the L-W value and the
results of $T_e$ and $R_{23}$ methods are slightly larger (a bit more
than 0.5~dex).  Model M2 shows instead a good agreement of the oxygen
abundances determined by diagnostic methods with both L-W and M-W
determinations.  In this case, the $T_e$ and $R_{23}$ methods can be
used for the determination of the oxygen abundance.

 Thus, we can conclude, that $T_e$- and $R_{23}$-methods have 
  limitations in two circumstances: $(i)$ in the low-density limit in nebular
  enviroments with complex element abundance distributions along sight
  line; $(ii)$ when an additional radiation field contributes to the
  stellar radiation, as e.g. the cooling radiation of a hot plasma
  like the superwind bubble.  It is clear that a more detailed and
  punctual analysis of the diagnostic methods can be done only once
  our models are able to reproduce more faithfully the real physical
  characteristics of galaxies.  Notwithstanding these limitations, we
  believe that the preformed test is reasonable - we try to analyze
  the galaxy with the ``eyes of an observer'' and try to recover the
  galactic properties from observational methods.  At some level what
  we ``get out'' should match what we ``put in''.  Moreover, since we
  use two different diagnostic methods, we expect at least some
  agreement between them.  The failure to match these expectations
  entitles us to conclude that diagnostic lines have limitations in
  low-density, chemically complex regions.

\begin{table}
{\small
\begin{center}
  \caption{ (12 + log O/H) values obtained for the ChDS model weighted
    by luminosity (L-W) and mass (M-W), respectively, and compared to
    abundances derived by means of the $T_e$ and the $R_{23}$
    diagnostic methods. The two lowermost lines give the (12 + log
    O/H) abundances for both analytical model distributions M1 and M2
    at 140 Myr, respectively.  }
\begin{tabular}{lcccccc}
\hline
Age,    & AP 	&\multicolumn{2}{c}{Weighted, from ChDS}&\multicolumn{2}{c}{Diagnostic methods}\\
Myr	& 	&	L-W   & 	M-W 		&	 $T_e$ &	 $R_{23}$(McCaugh)\\
\hline              
10	&Central&7.90 & 7.80 &  6.90 & 6.48 \\
10	&   1	&7.14 & 7.23 &  7.16 & 6.80 \\
10	&   2	&7.70 & 7.58 &  7.27 & 7.01 \\
10	&   3	&7.89 & 7.74 &  6.96 & 8.98 \\
10	&   4	&5.78 & 5.78 &  5.84 & 6.91 \\
10	&   5	&5.78 & 5.78 &  7.01 & 6.67 \\
\hline              
70	&Central&8.52 & 8.38 &  7.38 & 6.72 \\
70	&   1	&8.29 & 8.29 &  6.84 & 6.61 \\
70	&   2	&9.13 & 9.09 &  6.96 & 9.06 \\
70	&   3	&8.51 & 8.37 &  7.48 & 9.01 \\
70	&   4	&5.77 & 5.78 &  5.84 & 6.92 \\
70	&   5	&5.78 & 5.78 &  6.95 & 6.73 \\
\hline              
140	&Central&8.93 & 8.98 &  8.05 & 8.91 \\
140	&   1	&7.79 & 8.22 &  7.42 & 6.97 \\
140	&   2	&8.51 & 8.53 &  7.44 & 8.99 \\
140	&   3	&9.00 & 9.03 &  7.91 & 8.94 \\
140	&   4	&8.44 & 8.43 &  8.35 & 8.82 \\
140	&   5	&5.78 & 5.78 &  7.38 & 6.88 \\
\hline              
140	&   M1	&8.07 & 7.78 &  7.51 & 7.11 \\
140	&   M2	&8.21 & 7.79 &  8.10 & 7.96 \\
\hline              
\end{tabular}
\end{center}
\label{Tab:OxygenAbund}
}
\end{table}

\subsection{The empirical determination of the star formation rate in galaxies}
\label{subs:kennicutt}
\begin{figure}
\centering
\includegraphics[width=6cm,angle=-90]{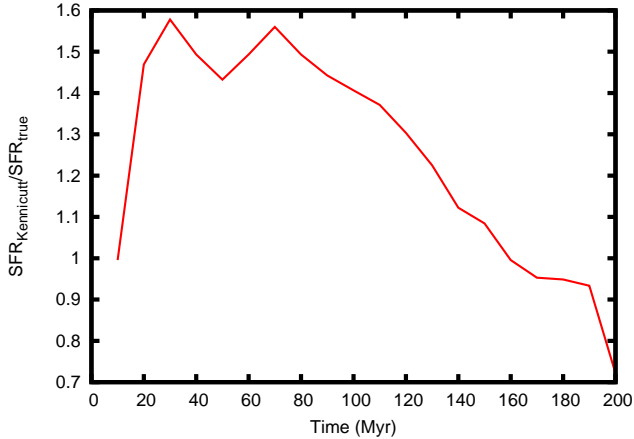}
\caption{The ratio of the star-formation rate determined from $L({\rm
    H\alpha})$ (see Eq. \ref{eq:kenn98}) and the 'true' rate, i.e.
  the rate which is used as input from our ChDS
  (2.67~$\cdot$~10$^{-2}$~M$_\odot$~yr$^{-1}$) of model M1.}
 \label{fig:halpha} 
\end{figure}

We conclude this section with an estimate of the reliability of
empirical indicators for the star formation rates in external galaxies.
One of the most commonly used proxy of SF in galaxies is the
\Ha emission.  In particular, the calibration of \citet{Kenn98}
\begin{equation}
{\rm SFR}\,({\rm M}_\odot {\rm yr}^{-1})=\frac{L(H\alpha)}{1.26 \times 
10^{41} \,{\rm ergs}\,{\rm s}^{-1}},
\label{eq:kenn98}
\end{equation}
\noindent
(based on the Salpeter IMF) is very widely used.  Since we are able to
calculate the \Ha luminosity by means of the MPhM code, we are
in the condition to compare the SFR based on $L({\rm H\alpha})$
with the 'true' SFR, i.e. the rate which has been used as input
for our ChDS (equal to 2.67~$\cdot$~10$^{-2}$~M$_\odot$~yr$^{-1}$, see
Sect. \ref{sec:ChDS}).  This comparison is shown in Fig.
\ref{fig:halpha}.

We can see from this figure that, at the beginning, there is a perfect
match between the \Ha-based and the 'true' SFR.  This reassures us
about the correctness of our procedures and the accuracy of the
Starburst99 code.  At later times, however, the \Ha luminosity
increases due to the contribution of the warm gas.  The Kennicutt SFR
estimates becomes thus inaccurate.  It overestimates the true SFR by
up to 60 \% at t=30 Myr. After 160 Myr, Lyc photons begin to leak out
of the galaxy (see Sect. \ref{sec:fesc}), thus the global \Ha
luminosity decreases and the relation Eq. \ref{eq:kenn98} starts
underestimating the true SF rate.  An accurate study of the status of
the gas in a galaxy is thus a fundamental prerequisite in order to
assess the reliability of the SF rate estimate based on \Ha emission.
 Given the model limitations, an agreement between model
  results and expectations within a factor of two should be considered
  as fair.  Still, we believe it is important to quantify these
  inconsistencies and to outline which physical processes can lead to
  inaccuracies in the Kennicutt's SFR determinations based on \Ha
  emission.

\section{The escape fraction of the ionizing photons}
\label{sec:fesc}
In this Section, we investigate how many ionizing photons escape the
galaxy volume.  These photons could play a fundamental role in the
reionization of the Universe \citep{Vanz12,KC14,Wise14}.

Let the escape fraction of ionizing photons from sector $i$ be defined
as follows
\begin{equation}
f_{esc}(i) = \frac{Q_{esc}(i)}{Q_{stars}(i)} = \frac{\int_{\nu_{1Ryd}}^\infty\frac{F_\nu(esc,i)\ d\nu}{h\nu}}{\int_{\nu_{1Ryd}}^\infty\frac{F_\nu(stars,i)\ d\nu}{h\nu}},
\end{equation}
\noindent
where $Q_{esc}(i)$ denotes the rate at which ionizing photons leave
sector $i$ through the outer edge (i.e. propagate above 2.3 kpc);
$Q_{stars}(i)$ is the rate of incident ionizing photons emitted by the
stars contained in sector $i$; $F_\nu(esc,i)$ and $F_\nu(stars,i)$ are
the corresponding fluxes of ionizing radiation.  It must be noted that
$F_\nu(stars)$ is not attenuated by the intervening gas; it is the
overall flux produced by all central stars.  In practice, we perform
an integration of both ionizing fluxes over the photon energy range
$(1-10)Ry$ for $F_\nu(stars)$, and $(1-180)Ry$ for $F_\nu(esc)$,
respectively.  This is due to the fact that $F_\nu(stars,i)$
(calculated, as mentioned above, using {\it Starburst99}) drops to
zero beyond $10Ryd$, but the escape radiation ($F_\nu(esc,i)$)
contains also the photons at higher energies, emitted by the hot
rarefied gas in the SWR.  Notice also that high-energy Lyc photons are
less absorbed, because the photoionization cross section decreases
with photon energy increasing.

The total escape fraction of the ionizing photons is defined as
\begin{equation}
f_{esc} = \frac{\sum_{i=1}^{20}Q_{esc}(i)}{\sum_{i=1}^{20}Q_{stars}(i)},
\end{equation}
\noindent
i.e. we simply sum up all the escape fractions, in all sectors.

\begin{figure}
\centering
\includegraphics[width=6cm,angle=-90]{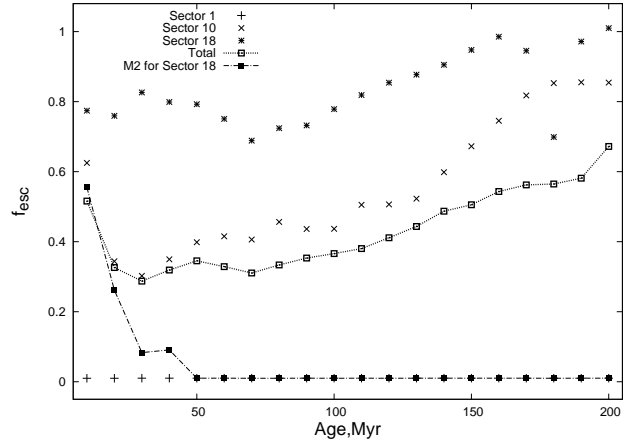}
\caption{Evolutionary dependences of the escape fraction 
$f_{esc}$ of ionizing photons for sectors 1, 10 and 18 as well as for total
  escape fraction (see text). Also, the evolution of $f_{esc}$ for the
  model M2 (Sector 18) is shown. The values of $f_{esc}$ for the M2
  model along Sectors 1 and 10 are $\approx$ 0.01 at all ages, thus
  have not been shown.}
\label{fig:Fesc}
\end{figure}

In Figure \ref{fig:Fesc} the evolution of $f_{esc}$ for sectors 1, 10,
and 18 as well as the total escape fraction is shown.  As expected,
lower values of $f_{esc}$ belong to low-number sectors, closer to the
disk of the galaxy.  The escape fraction of ionizing photons for the
M2 model are in general very low (of the order of 0.01).  Only the
escape fraction for sector 18 (close to the symmetry axis) is
significantly larger than 0, at least until 50 Myr.  This is thus the
only result we show for model M2 (see filled squares+dot-dashed curve
in Fig. \ref{fig:Fesc}).

We can notice from Figure \ref{fig:Fesc}, that $f_{esc}$ for Model M1
is on average 0.4. This value is slightly higher than the value found
in radiative hydrodynamical simulations \citep{KC14,Wise14}. On the
other hand, model M2 produces much lower values for $f_{esc}$, lower
than predicted by simulations and also lower than the values required
to ensure that DGs are significant for the reionization of the
Universe.  \citet{Fujita03} e.g. found that for DGs with SF
efficiencies, i.e. the gas fraction converted into stars above 6\%,
$f_{esc}$ can be larger than 0.2 and this value is large enough for a
significant contribution to the ionizing UV background.  Notice
however that the true SF efficiency in DGs could be smaller than 6\%.
In this case, DGs can not be the main responsible for the reionization
of the universe.  In the model L8F considered here, this SF efficiency
is about 13\%, thus our average $f_{esc}$ of $\sim$ 0.4 is consistent
with Fujita's calculations.

As we have shown above, the small scale TDS, characteristic of the M2
model, is necessary to reproduce the observed emission line spectra.
Of course, in the presence of many TDSs in the DG, the total $f_{esc}$
would be further reduced.  However, we shall not forget that our
calculations are based on the outward-only approximation and it is
possible that a fraction of the ionizing photons are emitted in
non-radial directions.  This would enhance the $f_{esc}$ in model M2.
It appears that some modifications in our modelling procedures (in
particular, the relaxation of the outward-only approximation) are
required in order to obtain $f_{esc}$ values closer to realistic ones
and less dependent on model assumptions.

\begin{figure}
\centering
\includegraphics[width=6cm,angle=-90]{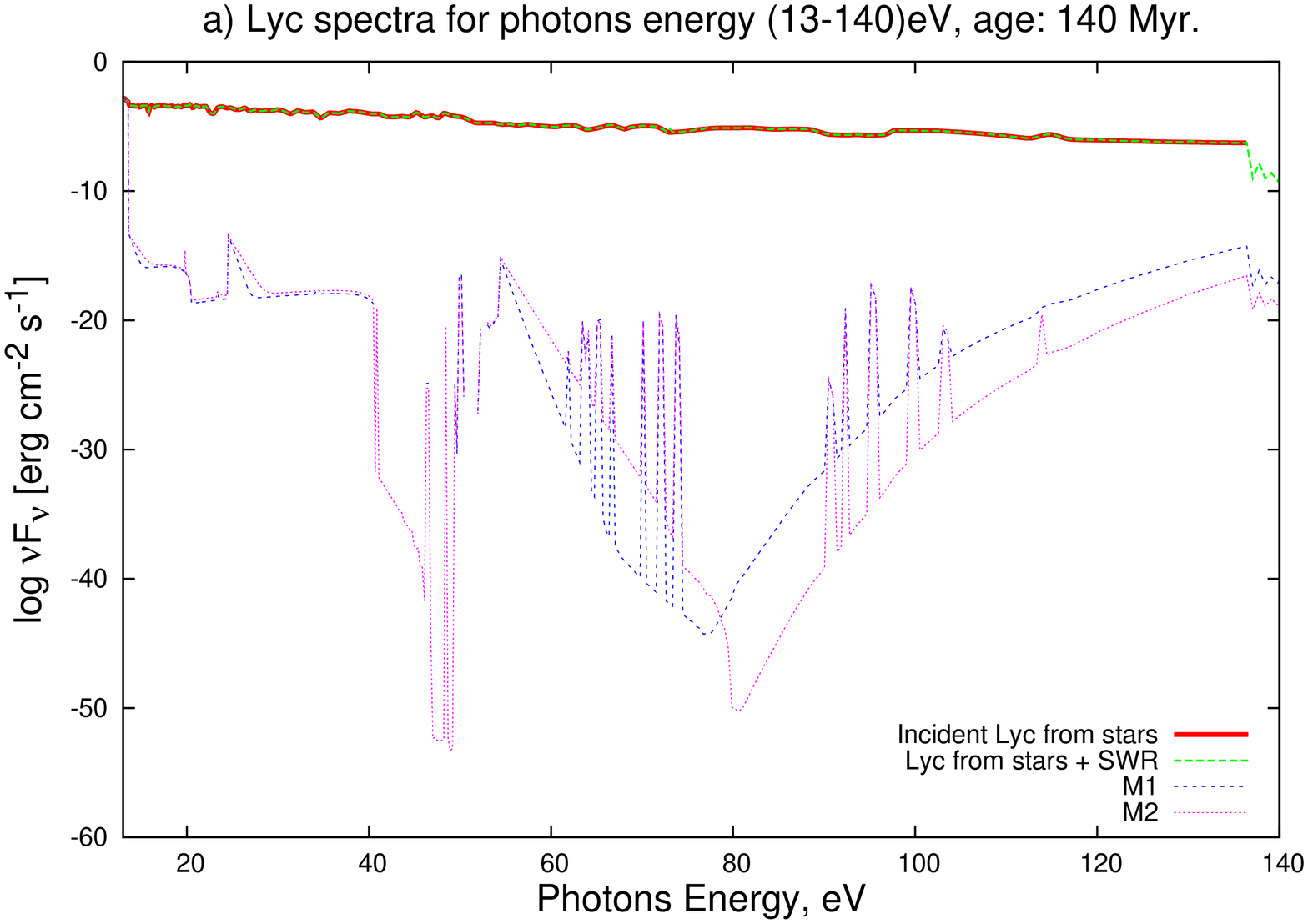}
\includegraphics[width=6cm,angle=-90]{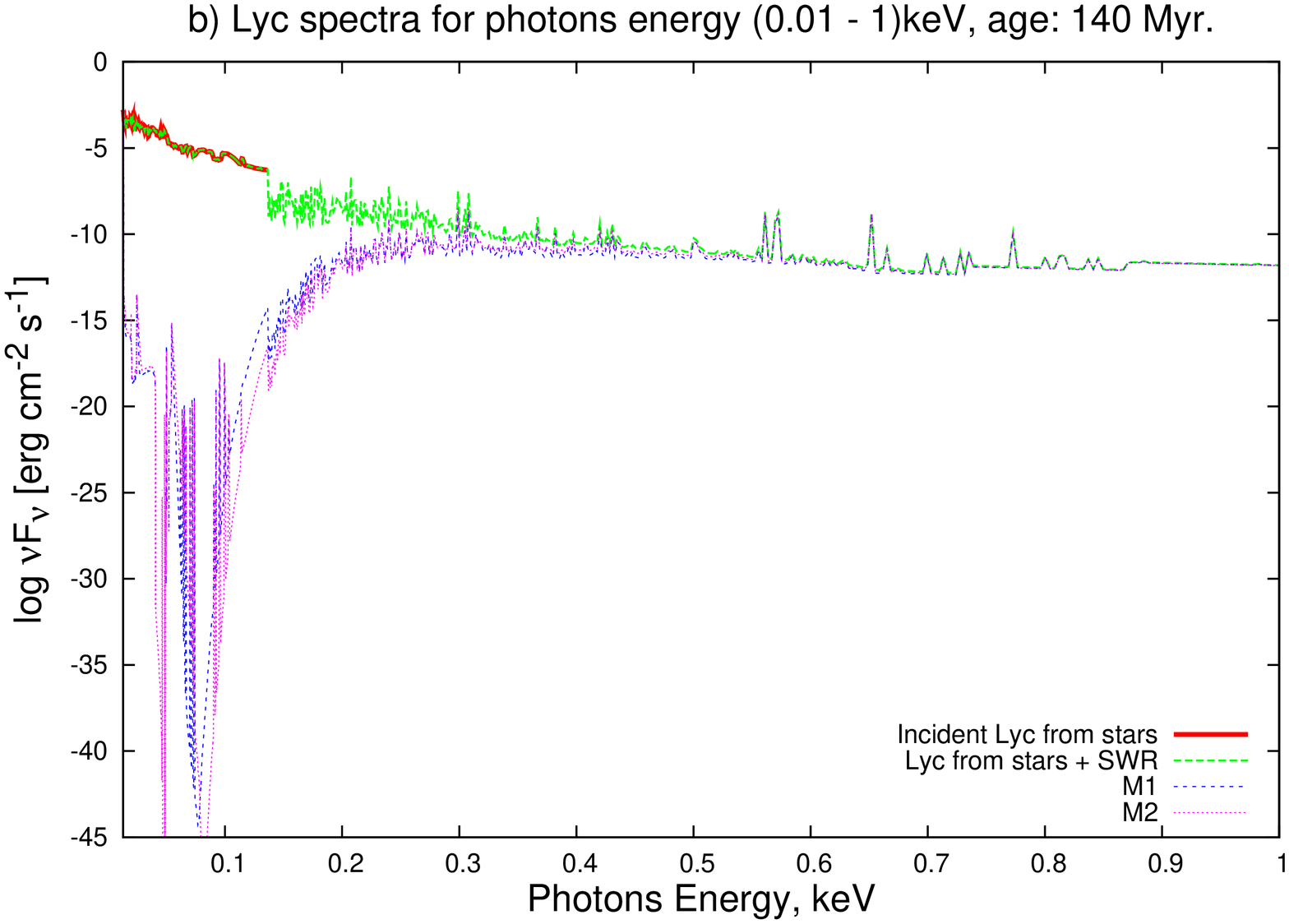}
\caption{Changes of the Lyc-spectrum from the stars during the
  transfer of ionizing radiation through the DG.  The upper panel
  refers to the energy range between 13 and 140 eV.  The lower panel
  extends up to 1 keV.  All fluxes are given at a distance of 2.3 kpc
  from the center of galaxy (at the outer edge of the modelling
  volume).}
\label{fig:LycTrans}
\end{figure}

We also studied the changes of the energy distribution in the ionizing
spectrum. For both M1 and M2 models, we show only results relative to
sector 1.  In Figure \ref{fig:LycTrans}a) we compare 1) the incident
Lyc-spectrum from central stars in the energy range $(13-140) eV$ for
both models; 2) the spectrum of the radiation escaping the SWR; 3) the
spectrum calculated at the outer edge of galaxy, predicted by model
M1.  The outer part of the galaxy absorbs very efficiently ionizing
quanta in this photon energy range, because of the high density in
Sector 1.  Therefore, an outer ionization front is present in this
sector.  The Lyc spectra at the outer edge are very weak.
Nevertheless, gaps in the Lyc spectra, mainly positioned at important
ionization potentials of ions, are clearly distinguishable.  Similar
gaps were predicted by optimized photoionization models of HII regions
in low-metallicity starburst galaxies \citep[see e.g.][]{Melekh2006,
  Melekh2007}.  As a result of these models, the optimal Lyc-spectra
for HII regions in the blue compact galaxies SBS~0940+544 and
SBS~0335-052 was found.  The gaps appear also in models of superwind
bubbles \citep{KM2013,KM2014}.

As mentioned above, the incident Lyc spectrum from stars drops to zero
at photon energies $\ge 136 eV$ (see Figures \ref{fig:LycTrans}a,b),
but the emission of warm-hot gas contributes photons at higher
energies.  The spectra in the range $(0.01-1) keV$ are shown in Figure
\ref{fig:LycTrans}b.

It can be seen that the density distribution in the outer part of
galaxy (from 'wall' up to outer edge of the modelling volume) shapes
the spectra up to photon energies of $(0.4-0.5) keV$. The spectrum at
larger photon energies is practically undisturbed, because of low the
optical thickness for these photons.  Therefore, the modelled spectrum
in this energy range can be compared with X-ray observations.

It must be also once again remarked that in our models we calculated
the diffuse component of the the ionizing radiation in outward only
approximation.  Probably, the diffuse ionizing radiation does not play
a crucial role in the formation of the global structure of diffuse
ionized gas.  Nevertheless, as we have shown above, important emission
lines can be partly formed in low-ionization zones of small spatial
scale.  Existing 3D codes \citep[such as Mocassin;][]{Ercolano} can
calculate the diffuse ionizing field in regions shadowed by TDS, but
they need a very large amount of computing time.  Moreover, Mocassin
is based on the statistical analysis of random propagation of photon
packages.  We think that for the problem at hand a deterministic
approach, based on the detailed calculation of diffuse radiation
propagation, is more useful.  Therefore, we are developing an
extension of the MPhM, able to follow in detail the diffuse ionizing
radiation field (i.e. we will relax the outward-only approximation).
Such method can be very useful for the correct investigation of the
ionization structure in the shadow regions beyond the TDS.  As
mentioned above, this method can also help simulate the leakage of
ionizing photons from a galaxy in a more realistic way.

\section{Discussion} 

In this paper we have described a way to combine detailed ChDS of DGs
with a state-of-the-art MPhM, to study the emission properties of
star-forming galaxies and compare them with observations.  

We have performed two types of analysis of the obtained MPhM results:
$(i)$ {\it model predictions} and $(ii)$ {\it consistency checks}.
Important emission line ratios have been calculated and compared with
observations of a DG galaxy with structural parameters similar to the
simulated one (see Sect. \ref{subs:intratios}).  The calculated escape
fraction of ionizing photons $f_{esc}$ is calculated, too, and
compared with results of radiative transfer hydrodynamical simulations
(Sect. \ref{sec:fesc}).  The results of these calculations are briefly
summarized and discussed in Sect. \ref{disc:pred}.  With consistency
checks we mean instead tests in which some galactic properties are
known because of the results of the ChDSs.  The same galactic
properties are re-derived as an observer would do, based on the MPhM
results.  The analysis of the oxygen abundances (Sect.
\ref{subs:abundances}) and of the empirical determinations of the SFR
in galaxies (Sect. \ref{subs:kennicutt}) belong to this category.  The
results of these two tests are briefly discussed below (Sect.
\ref{disc:cons}).  We shall not forget that our analysis is based on
some simplifying hypotheses and there are potentially important
physical processes (e.g. the dust reprocessing) that have not been
considered.  We will briefly discuss these limitations in Sect.
\ref{disc:limitations}.  Finally, some overall conclusions are drawn
in Sect. \ref{disc:conclusion}.

\subsection{Model predictions}
\label{disc:pred}

We have calculated the intensities of
observationally relevant emission lines both integrating along angular
sectors (Sect. \ref{subs:sectors}) and across apertures (Sect.
\ref{subs:APs}).  
Some line intensities and intensity ratios, such as 
[OIII]$\lambda5007$/H$\beta$ and H$\alpha$/H$\beta$, 
nicely fit into the range of observed intensity
ratios (see Table 2) whereas some others do not.

In particular, calculated [OII] line intensities are very weak. It
should be also remarked here that the average final oxygen abundance
of the model L8F is 12+log(O/H)=7.84 (see RH13).
We have also calculated model M1 at a higher metallicity,
corresponding to the oxygen abundance 12+$\log$O/H=8.2.  For such
model, the calculated [OII] line intensities are still weak in
comparison with observations.  This is due to the fact that the
supershell created by the stellar feedback in ChDSs is not dense
enough to block O-ionizing photons.  O$^+$ is thus very scarce (see
Fig. \ref{fig:OxygenM}) and this clearly implies very few O$^+$
transitions.  The inclusion of artificial density enhancements close
to the supershell (model M2) helps creating a shell of singly ionized
oxygen.  For model M2, thus, also intensity ratios such as
[OII]$\lambda3727$/$H\beta$ are in agreement with observations.

We will provide a comprehensive study on the models like M1 and M2
at different metallicities in a future paper.

We have also calculated, for each angular sector, the 
escape fraction of ionizing photons $f_{esc}$.  This fraction is
generally very high along directions close to the symmetry axis, i.e.
towards the poles, because the superbubble propagates preferentially
along these directions due to the very steep pressure gradients.  The
global $f_{esc}$ ranges from $\sim$ 0.3 to $\sim$ 0.6, with an average
of about 0.4.  Results of radiative hydrodynamical simulations of DGs
generally show lower values of $f_{esc}$, although the results
crucially depend on the details of the feedback scheme and on the
adopted star formation efficiencies \citep[see e.g.][]{Fujita03}.
Escape fractions calculated for model M2 are instead generally low.
They can be of the order of 0.3--0.5 only along directions close to
the symmetry axis and only for a limited amount of time, otherwise
their values are of the order of 0.01.  With such low values of
$f_{esc}$ one can not expect that DGs significantly contribute to the
reionization of the universe as concluded by \citet{Vanz12}. 
 In M2 models it was assumed that the TDS has the
 same density in all sectors. Of course this is a just a first-order
 approximation. TDS in different sectors would be characterized by
 different density values; in some sectors TDS might be even absent.
 Therefore, in reality, the Lyc photon escape should be higher in the
 vertical direction. Thus, evidently, more accurate calculations are
necessary to address more in detail the issue of the escape fraction
(see also Sect.  \ref{disc:limitations}).

\subsection{Consistency checks}
\label{disc:cons}

The employed ChDSs follow in detail the evolution
in space and time of important chemical elements, in particular
oxygen.  Based on the chemical composition of different regions of the
galaxy, the MPhM calculates intensity ratios, as extensively described
in the previous sections.  On the other hand, these intensity ratios
can be measured in real galaxies and are the principal way observers
can infer the chemical composition of galaxies.
We study the results of our simulations as an observer would do and
derive the oxygen abundances based on calculated intensity ratios.  As
explained above, we use the $T_e$ and $R_{23}$ diagnostic methods.

In an ideal world, the abundances derived by means of intensity
ratios should match the 'real' ones, i.e. here the ones obtained
from the ChDSs.  This turns out to be the case only for some apertures
and for some intervals of time (see Table 3).  For other apertures
and/or other intervals of time, the differences are significant.
These disagreements are due to two reasons: $(i)$ the distribution of
oxygen is very inhomogeneous and these inhomogeneities cannot be
properly addressed by the applied diagnostic methods.  These
diagnostic methods work well in relatively uniform regions, but fail
in the presence of very inhomogeneously distributed chemical elements.
$(ii)$ Diagnostic methods commonly assume 
that most of the oxygen is either singly or/and doubly ionized.  This
is not the case in model M1, as we have shown in Fig.
\ref{fig:OxygenM}.  For this reason, the adoption of unresolved
density enhancements in thin shells (model M2) helps in obtaining
'observed' ionization stages of oxygen close to the 'real' ones.

In spite of that, our results indicate that the diagnostic methods should
be used with a 'pinch of salt' to say the less.  For galaxies in which
integral field spectroscopy is available, the difficulties of the
diagnostic methods are reduced, as every chunk of the galaxy encloses
a different abundance but the small-scale abundance inhomogeneities
affect the integrated spectral lines much less. Notice also
that \citet{Leroy12} report of 1~kpc-scale variations
and uncertainties of observed SFR indicators that are inherently e.g.
caused by age effects and by intrinsic scatters of the indicators.

The \Ha emission from galaxies is extensively applied to derive their
SFRs locally and globally \citep[see][and references therein]{Kenn98},
because it is in fact assumed to be the signpost of (short-living)
massive stars. For an assumed IMF the number of massive stars is,
therefore, correlated with the SFR.
As we have shown (Fig.  \ref{fig:halpha}) the total \Ha emission is up
to $\sim$ 60\% larger than the one expect from the stellar ionizing
flux of our model. If one takes into account that, in addition,
ionizing photons are also lost by the leakage from the galaxy after a
few tens of Myr (Fig.  \ref{fig:Fesc}) and that the \Ha-SFR relation
by \citet{Cal07} implies an even lower proportionality constant
between SFR and \Ha luminosity, one can conclude that the additional
ionization by cooling photons from the warm-hot gas is almost equal to
the stellar contribution.  At later times, the \Ha emission reduces,
partly, due to the increasing amount of leaking ionizing photons.  In
this phase, an \Ha-based determination of the SFR would underestimate
the real galactic SFR (see again Fig. \ref{fig:halpha}).  
However, given the uncertainties associated to this SFR indicator, a
factor of two agreement can be considered as fair.

Actually, it has been already reported in the literature that \Ha
could be a poor SFR indicator in DGs. In particular, it has been
noticed that the \Ha-based SFR estimates differ from the SFR estimates
based on the UV emission at SFRs smaller that $\sim$ 0.01 M$_\odot$
yr$^{-1}$ and the discrepancy can be larger than a factor of 10 at
SFR=10$^{-4}$ M$_\odot$ yr$^{-1}$ \citep{Lee09}.  This discrepancy has
been associated to a top-light IMF in DGs with very mild SFRs
\citep{Pflamm09}. In such galaxies, there is some star formation but
not enough generation of the very massive stars able to produce \Ha
emission. Notice, however, that our SFR (nominally
2.67~$\cdot$~10$^{-2}$~M$_\odot$~yr$^{-1}$) is not extremely low.  In
this SFR range the effect of the top-light IMF should not be very
significant.  Notice also that the leakage of \Ha photons can be
significant for our model and it is surely even more significant for
galaxies with lower masses, where galactic winds are more prominent
(see RH13). It is thus reasonable to assume that the discrepancy
between \Ha-based and UV-based SFRs noticed by \citet{Lee09} also
depends on $f_{esc}$.  Notice however also that, as \citet{Leroy12}
point out, Starburst99 simulations of an evolving single stellar
population imply an intrinsic scatter of almost 0.3 dex in \Ha-based
SFR estimates and about 0.5 dex in FUV-based determinations. We will
add these issues in detail in a follow-up paper.

\subsection{Model limitations}
\label{disc:limitations}

We have already outlined three limitations of the study presented in
this paper, namely:
\begin{itemize}
\item The MPhM is applied in a post-processing step, thus the energy
  due to radiative processes does not feed back on the gas and does
  not affect the hydrodynamics.
\item The gas densities, obtained from the ChDSs, are always below a
  few particles per cm$^{3}$, i.e. they never reach the peak densities
  of typical HII regions.
\item The MPhM calculations are based on the outward-only
  approximation (the only one available with CLOUDY)
\end{itemize}
We do not plan to relax the first simplification because this would
require computationally expensive radiative hydrodynamics simulations.
These calculations do not allow to explore a wide parameter space and,
moreover, simplifying assumptions about the radiation field and the
propagation of radiation need to be made.  Notwithstanding the
limitations, we think that our approach is reasonable because it is
computationally inexpensive and the radiation transfer equations can
be calculated with great accuracy.  We do plan to relax the last two
simplifications though.

The second limitation can be relaxed by using a multi-phase ISM
treatment, that takes into account (in the same computational cell) a
warm-hot, diluted phase, and a colder, denser phase, that can reach
the peak densities of typical HII regions and more.  Notice
also that, for the galaxy NGC1569, \citet{West07,West08}
concluded from \Ha line decompositions that the light profile is
composed of a bright, narrow feature originating within a narrow
region of $\sim$700$\times$500~pc in size, roughly centred on the
bright super star cluster A, and of an underlying broad component
emerging from turbulent mixing layers on the surfaces the clouds which
are embedded into the hot, fast-flowing winds from the young star
clusters and experience evaporation and/or ablation of material. A new
3D chemo-dynamical code (cdFLASH) is already available (Mitchell et
al., in preparation) and needs to be applied to the problem at hand.

The last limitation (the outward-only approximation) will be relaxed
once a new CLOUDY wrapper (currently under development) will be
available.

It is, however, worth mentioning that the model presented here lacks
another potentially relevant ingredient: dust.  Dust can increase the
opacities of many lines along all directions, thus affecting our
results significantly.  The reason why dust has not been included so
far is not technical: the MPhM wrapper and the underlying CLOUDY code
can already treat dust.  We have decided to neglect dust only for the
purpose of clarity and in order to avoid cluttering of the (already
long) text.  We have however already performed a pilot study with
dust.  Assuming a gas-to-dust ratio of 1400
(ten times larger than the Milky Way value, typical for DGs)
we find very little differences between the results
of a dusty model and the ones presented in this paper.  However, for
smaller gas-to-dust ratios the effect of dust can be very significant.
We will explore the dust effect in a short follow-up paper.

\section{Conclusions}
\label{disc:conclusion}

The main conclusions of the present study can be summarized as 
follows:
\begin{itemize}
\item Our combined ChDS-MPhM approach is able to produce intensity
  ratios of important emission lines which are in agreement with the
  observations.  This is not the case for O$^+$ lines, which are
  severely underestimated.  The reason is that the densities are 
  on average low and most of oxygen in the model galaxy is in high 
  ionization stages.  
\item An artificial density enhancement (simulating an
  unresolved thin shell) brings also [OII] lines in agreement with
  observations. 
\item We obtain typical escape fractions of ionizing photons
  ($f_{esc}$) of the order of 40\%.  This escape fraction reduces to a
  few per cent if we consider the same density enhancements described
  above.
\item Diagnostic methods ($T_e$ and $R_{23}$) are generally not able
  to reproduce the real abundance of heavy elements, obtained by means
  of the chemo-dynamical simulations.  This is due in part to the
  inherent limitations in the diagnostic methods, in part to the fact
  that in our simulation very little oxygen is neutral or singly
  ionized.
\item We have shown that the Kennicutt's estimate of the SFR in
  galaxies based on the \Ha emission is inaccurate.  Part of the
  inaccuracy is due to the significant contribution of the warm-hot
  gas to the \Ha emission.  Part is instead due to the fact that
  the leakage of ionizing photons out of a typical starbursting galaxy
  can be significant and that naturally affects the global \Ha
  emission.
\end{itemize}


\section*{Acknowledgements}
B. Melekh thanks the financial support of Austrian Exchange Service
(OeAD) during his visits to the Dept. of Astrophysics at University
of Vienna and the hospitality of the Institute. 

The authors thank an anonymous referee for a careful reading of the
manuscript and useful comments which help clarifying the paper
content.

\end{document}